\documentclass[twocolumn,showpacs,aps,prd]{revtex4}
\usepackage{xspace}
\usepackage{graphicx}
\usepackage{dcolumn}
\usepackage{amsmath}
\usepackage{epsfig}
\usepackage{afterpage}
 
\include{babarsym}

\def\taufiveh       {\ensuremath{\taum \! \rightarrow  3h^- 2h^+ \nut}\xspace }

\def\taufivepi      {\ensuremath{\taum \! \rightarrow  3\pim  2\pip   \nut}\xspace }
\def\taufivepipiz   {\ensuremath{\taum \! \rightarrow  3\pim 2\pip \piz \nut}\xspace }
\def\taupppzzz      {\ensuremath{\taum \! \rightarrow  2\pim \pip 3\piz  \nut}\xspace }
\def\taupzzzzz      {\ensuremath{\taum \! \rightarrow  \pim 5\piz  \nut}\xspace }

\def\taupppe        {\ensuremath{\taum \! \rightarrow  2\pim \pip \eta  \nut}\xspace }

\def\taupzzeta      {\ensuremath{\taum \! \rightarrow  \pim 2\piz \eta  \nut}\xspace }
\def\tauthreepieta  {\ensuremath{\taum \! \rightarrow  (3\pi)^- \eta  \nut}\xspace }

\def\tauetaprime    {\ensuremath{\taum \! \rightarrow  \pim \etapr (958) \nut}\xspace }
\def\taupietaprimepiz{\ensuremath{\taum \! \rightarrow  \pim \piz \etapr (958) \nut}\xspace}
\def\tauketaprime   {\ensuremath{\taum \! \rightarrow  \Km \etapr(958) \nut }\xspace }

\def\taufpi         {\ensuremath{\taum \! \rightarrow \pim f_1 \nut}\xspace }

\def\taupppo        {\ensuremath{\taum \! \rightarrow  2\pim \pip \omega  \nut}\xspace }

\def\taupzzomega    {\ensuremath{\taum \! \rightarrow  \pim 2\piz \omega  \nut}\xspace }
\def\tauthreepiomega{\ensuremath{\taum \! \rightarrow  (3\pi)^- \omega  \nut}\xspace }

\def\taukfourpi     {\ensuremath{\taum \! \rightarrow K^- 2\pim  2\pip   \nut}\xspace }
\def\taukoppfourpi  {\ensuremath{\taum \! \rightarrow K^+ 3\pim  \pip   \nut}\xspace }
\def\taukkppp       {\ensuremath{\taum \! \rightarrow K^- K^+ 2\pim  \pip   \nut}\xspace }

\def\taukfourpipiz  {\ensuremath{\taum \! \rightarrow K^- 2\pim  2\pip \piz \nut}\xspace }
\def\taukoppfourpipiz{\ensuremath{\taum \!\rightarrow K^+ 3\pim   \pip \piz \nut}\xspace }

\def\etappp         {\ensuremath{\eta \! \rightarrow \pip \pim \piz}\xspace }
\def\etagg          {\ensuremath{\eta \! \rightarrow \gamma \gamma}\xspace }
\def\etapiz         {\ensuremath{\eta \! \rightarrow 3\piz}\xspace }

\def\fpppp          {\ensuremath{f_1   \! \rightarrow  2\pip 2\pim}\xspace }
\def\feta           {\ensuremath{f_1   \! \rightarrow  \pip \pim \eta}\xspace }
\def\fak            {\ensuremath{f_1   \! \rightarrow  a_0^- \pip}\xspace }

\def\azeta          {\ensuremath{a_0^-(980)  \! \rightarrow  \pim \eta}\xspace }

\def\etap           {\ensuremath{\etapr \! \rightarrow \pim \pip \eta}\xspace }

\def\omegappp       {\ensuremath{\omega \! \rightarrow \pim \pip \piz}\xspace }
\def\omegappg       {\ensuremath{\omega \! \rightarrow \pim \pip \gamma}\xspace }

\def\fpdatax{68985 \pm 263}

\def\fpseleffp{(7.98 \pm 0.02)\%}
\def\fpbfpi{4441 \pm 370}

\def\fpbtot{16233 \pm 835}
\def\fpbtau{10621 \pm 719}
\def\fpbqq {1171 \pm 205}
\def\fpeff{0.2}
\def\fplumi{1.0}
\def\fptk{3.8}
\def\fplep{1.6}
\def\fpbk{1.6}
\def\fppi{2.5}
\def\fptot{5.2}
\def\fpbf{(7.68 \pm 0.04 \pm 0.40)}
\def\fpbfx{7.68 \pm 0.04 \pm 0.40}
\def\fpbfinc{(8.33 \pm 0.04 \pm 0.43)}

\def\ffadnevt{3722\pm 222}
\def\ffadchisq{77}
\def\ffadndf{62}

\def\ffeffp{(8.3\pm 0.1)\%}
\def\ffbr{(4.73\pm 0.28\pm 0.45) \times 10^{-4} }

\def\ffbrex{(5.20\pm 0.31\pm 0.37) \times 10^{-5} }
\def\ffbrexx{0.520\pm 0.031\pm 0.037}
\def\ffssca{3.8}
\def\ffsscb{2.5}
\def\ffsscc{1.6}
\def\ffsscd{1.0}
\def\ffsua{5.0}

\def\ffsuc{0.6}

\def\ffssex{7.0}

\def\sixnevts{7296\pm 85}

\def\sixEffp{(3.71\pm 0.03)\%}
\def\sixbr{( 3.6\pm  0.3\pm  0.9) \times 10^{-5} }
\def\sixbrx{0.36\pm 0.03\pm 0.09}
\def\sixbrsum{(1.65\pm 0.05\pm 0.09) \times 10^{-4} }
\def\sixbriso{(1.11\pm 0.04\pm 0.09) \times 10^{-4} }

\def\sixbkgds{4458\pm 244}

\def\sixbkgd{6132\pm 267}
\def\sixbkgdt{1315\pm 100}
\def\sixbkgdq{359\pm 37}
\def\sixsa{1.0}
\def\sixsb{3.8}
\def\sixsc{1.6}
\def\sixsd{2.5}
\def\sixse{3.0}

\def\sixsh{0.9}
\def\sixsi{22.9}
\def\sixsyst{23.7}

\def\fonechisq{61}
\def\fonendf{55}
\def\fonenevt{731\pm 62}

\def\fgeffp{(2.97\pm 0.05)\%}

\def\fonebrbx{1.26\pm 0.11\pm 0.08 }

\def\fgssa{3.8}
\def\fgssb{3.0}
\def\fgssc{1.6}
\def\fgssd{2.5}
\def\fgsse{1.0}
\def\fgssf{1.2}
\def\fgssg{1.6}
\def\fgss{6.1}

\def\etachisq{60}
\def\etandf{52}
\def\etanevt{1440\pm 68}

\def\etaeffp{(2.97\pm 0.12)}

\def\etanqq{65\pm 38}

\def\etabrx{2.37\pm 0.12\pm 0.18}

\def\etasa{3.8}
\def\etasb{3.0}
\def\etasc{2.5}
\def\etasd{1.6}
\def\etase{1.8}
\def\etasf{1.0}
\def\etasg{2.8}
\def\etash{4.0}
\def\etass{7.4}

\def\omegachisq{55}
\def\omegandf{44}
\def\omeganevt{2372\pm 94}

\def\omeffp{(3.27\pm 0.03)\%}
\def\omnqq{257\pm 71}

\def\ombrb{( 8.4\pm  0.4\pm  0.6) \times 10^{-5} }
\def\ombrbx{8.4\pm 0.4\pm 0.6 }
\def\omsa{3.0}
\def\omsb{2.5}
\def\omsc{1.6}
\def\omsd{3.8}
\def\omse{1.0}
\def\omsf{0.8}
\def\omsg{3.4}
\def\omsh{0.8}
\def\oms{6.8}

\def\sccxndata{44\pm 11}
\def\sccxnmc{45\pm 12}

\def\sccxeffp{(3.25\pm 0.15)\%}

\def\sccxblimitb{9.0 \times 10^{-6} }
\def\sccxbrcx{-0.4\pm  3.9\pm  4.3 }

\def\sumpietaprimelimitb{4.0\times 10^{-6}}

\def\ketapeffx{(3.09\pm 0.04)\%}
\def\ketapdata{15\pm 4}
\def\ketapbkgd{11\pm 3}

\def\ketapbra{1.6\pm 1.4\pm 1.2}
\def\ketaplimit{4.2\times 10^{-6} }

\def\kfoureffx{(7.9\pm 0.1)\%}
\def\kfourdata{1328\pm 36}

\def\kfourbrx{0.6\pm 0.5\pm 1.1}

\def\kfournewlimit{2.4\times 10^{-6} }

\def\kfourtotal{1284\pm 72}

\def\koppfourdata{1999\pm 45}

\def\koppfourbrx{1.6\pm 0.6\pm 2.4}
\def\koppfourlimit{5.0\times 10^{-6} }

\def\koppfourbc{1890\pm 163}

\def\kfiveeffx{(2.9\pm 0.1)\%}
\def\kfivedata{112\pm 11}

\def\kfivebrx{1.1\pm 0.4\pm 0.4}
\def\kfivelimit{1.9\times 10^{-6} }
\def\kfivenewlimit{1.9\times 10^{-6} }

\def\kfivetotal{84\pm 10}

\def\koppfiveeffx{(2.9\pm 0.1)\%}
\def\koppfivedata{154\pm 12}

\def\koppfivebrx{-0.6\pm 0.5\pm 0.7}
\def\koppfivelimit{8\times 10^{-7} }

\def\koppfivebc{170\pm 16}

\def\kkfoureffx{(6.7\pm 0.1)\%}
\def\kkfourdata{32\pm 6}

\def\kkfourbrx{0.30\pm 0.10\pm 0.07}
\def\kkfourlimit{4.5\times 10^{-7} }

\def\kkfourbc{15\pm 4}

\def\EtaEffGam{(3.83 \pm 0.11)\%}
\def\NsigEtaGam{2887 \pm 103}

\def\EtaDataChiGam{107/76}

\def\NqqEtaMCGam{131 \pm 29}

\def\tauetaBRGamCorrNoBra{2.10\pm0.09\pm0.13}

\def\EtaGammaEtaQQSyst{1.0}
\def\EtaGammaEtaEffSyst{3.0}
\def\EtaGammaTrackingSyst{2.7}

\def\EtaGammaLumiSyst{1.0}
\def\EtaGammaPIDSyst{1.6}
\def\EtaGammaBgammaSyst{1.0}
\def\EtaGammaPiPIDSyst{1.5}

\def\EtaGammaSystTotal{6.3}

\def\fEffGam{(3.75 \pm 0.04)\%}
\def\NsigfGam{1605 \pm 94}

\def\fDataChiGam{50/43}

\def\taufBRnofetaGamNoBra{1.25\pm0.08\pm0.07}

\def\FoneEtaGammaEtaEffSyst{3.0}
\def\FoneEtaGammaTrackingSyst{2.7}
\def\FoneEtaGammaMCEffSyst{1.1}
\def\FoneEtaGammaLumiSyst{1.0}
\def\FoneEtaGammaPIDSyst{1.6}
\def\FoneEtaGammaBgammaSyst{0.7}

\def\FoneEtaGammaPiPIDSyst{1.5}
\def\FoneEtaGammaModeling{2.7}
\def\FoneEtaGammaSystTotal{5.6}

\def\EtaEffPiz{(0.42 \pm 0.01)\%}
\def\NsigEtaPiz{315 \pm 34}

\def\EtaDataChiPiz{31/34}

\def\NqqEtaMCPiz{13 \pm 7}

\def\BRThreePizNoBra{2.54 \pm 0.27 \pm 0.25}

\def\EtaThreePiZeroEtaEffSyst{9.0}
\def\EtaThreePiZeroTrackingSyst{2.7}
\def\EtaThreePiZeroMCEffSyst{2.8}
\def\EtaThreePiZeroLumiSyst{1.0}
\def\EtaThreePiZeroPIDSyst{1.6}
\def\EtaThreePiZeroBpizSyst{0.9}
\def\EtaThreePiZeroPiPIDSyst{1.5}
\def\EtaThreePiZeroBkgSyst{2.3}
\def\EtaThreePiZeroSystTotal{10}

\def\fEffPiz{(0.53 \pm 0.06)\%}
\def\NsigfPiz{197 \pm 59}

\def\fDataChiPiz{39/43}

\def\taufBRnofetaPizNoBra{1.33\pm0.39\pm0.20}

\def\FoneEtaThreePizEtaEffSyst{9.0}
\def\FoneEtaThreePizTrackingSyst{2.7}
\def\FoneEtaThreePizMCEffSyst{11}
\def\FoneEtaThreePizLumiSyst{1.0}
\def\FoneEtaThreePizPIDSyst{1.6}
\def\FoneEtaThreePizBpizSyst{0.9}

\def\FoneEtaThreePizPiPIDSyst{1.5}

\def\FoneEtaThreePizSystTotal{15}

\def\NsigEtaPiTwoPiz{381 \pm 45}

\def\EtaEffPiTwoPiz{(0.75 \pm 0.02)\%}

\def\NbkgEtaPiTwoPiz{83 \pm 12}

\def\EtaDataChiPiTwoPiz{95/75}

\def\tauetaCorrBRPiTwoPiz{(2.01\pm0.34\pm0.22)\times10^{-4}}

\def\tauetaCorrBRPiTwoPizNoBra{2.01\pm0.34\pm0.22}

\def\EtaPiTwoPiZeroSystErrLepPID{1.6}
\def\EtaPiTwoPiZeroEtaEffSyst{9.0}
\def\EtaPiTwoPiZeroTrackingSyst{2.7}
\def\EtaPiTwoPiZeroMCEffSyst{2.7}
\def\EtaPiTwoPiZeroLumiSyst{1.0}

\def\EtaPiTwoPiZeroBetapipipizSyst{1.8}
\def\EtaPiTwoPiZeroPiPIDSyst{1.5}
\def\EtaPiTwoPiZeroBkgSyst{4.0}
\def\EtaPiTwoPiZeroSystTotal{11}

\def\OmegaEffPiTwoPiz{(0.75 \pm 0.01)\%}
\def\NsigOmegaPiTwoPiz{1135 \pm 70}

\def\OmegaDataChiPiTwoPiz{42/44}

\def\OmegaNbkgTotal{709\pm59}

\def\BROmegaTwoPiPizCorr{(7.3 \pm 1.2 \pm 1.2)\times10^{-5}}

\def\BROmegaTwoPiPizCorrNoBra{7.3 \pm 1.2 \pm 1.2}

\def\OmegaPiTwoPiZeroEtaEffSyst{9.0}
\def\OmegaPiTwoPiZeroTrackingSyst{2.7}
\def\OmegaPiTwoPiZeroMCEffSyst{1.8}
\def\OmegaPiTwoPiZeroLumiSyst{1.0}
\def\OmegaPiTwoPiZeroPIDSyst{1.6}
\def\OmegaPiTwoPiZeroBomegapipipizSyst{0.8}
\def\OmegaPiTwoPiZeroPiPIDSyst{1.5}
\def\OmegaPiTwoPiZeroBkgSyst{14}
\def\OmegaPiTwoPiZeroSystTotal{17}

\def\NsigIncThreePiz{4094 \pm 64}

\def\EffPSIncThreePiz{(0.88 \pm 0.01)\%}
\def\NResMCIncThreePiz{1763 \pm 222}
\def\NQQIncThreePiz{573 \pm 50}

\def\NOthertSum{1681 \pm 44}
\def\NBkgPSIncThreePiz{4017 \pm 232}

\def\BRPSIncThreePiz{(1.0 \pm 0.8 \pm 3.0)\times10^{-5}}
\def\CLIncThreePiThreePiz{5.8\times10^{-5}}
\def\BRTotalIncThreePiz{(2.07\pm0.18\pm0.37)\times10^{-4}}
\def\BRPSIncThreePizNoEtaorFone{(16.9 \pm 0.8 \pm 4.3)\times10^{-5}}

\def\BRPSIncThreePizNoBra{0.10 \pm 0.08 \pm 0.30}

\def\PrimeGGEff{(4.06 \pm 0.35)\%}
\def\PrimeGGNsig{40 \pm 22}

\def\PrimeGGNqq{58 \pm 12}

\def\PrimeGGBRNoBra{-2.8\pm3.5\pm1.9}
\def\PrimeGGCL{5.2\times10^{-6}}

\def\PrimeThreePizEff{(0.96 \pm 0.10)\%}
\def\PrimeThreePizNsig{12 \pm 10}

\def\PrimeThreePizNqq{14 \pm 4}

\def\PrimeThreePizCL{1.3\times10^{-5}}

\def\PrimeThreePizBRNoBra{-1.8\pm8.1\pm3.3}

\def\EffEtaPrimeKaon{(3.47 \pm 0.03)\%}
\def\NdatEtaPrimeKaon{6\pm7}
\def\NqqEtaPrimeKaon{3\pm4}

\def\CLCorrEtaPrimeKaon{2.4\times10^{-6}}
\def\BRCorrEtaPrimeKaonNoBra{0.5\pm1.3\pm0.4}

\def\CLCorrEtaPrimePiz{1.4\times10^{-5}}

\def\EffEtaPrimePiz{(1.58\pm0.02)\%}
\def\NdatEtaPrimePiz{24\pm10}
\def\NqqEtaPrimePiz{5\pm7}

\def\BRCorrEtaPrimePizNoBra{7.8\pm4.1\pm1.7}

\def\pietappizxeff{(1.00\pm 0.03)\%}
\def\pietappizxdata{5\pm 6}
\def\pietappizxbkgd{5\pm 8}

\def\pietappizxbra{0.0\pm 7.6\pm 9.3}
\def\pietappizxlimit{1.9\times 10^{-5} }

\def\pietaprimepizlimit{1.2 \times 10^{-5} }
\def\ketaprimelimit{2.4 \times 10^{-6} }

\def\sumetabavg{(2.25\pm 0.07\pm 0.12) \times 10^{-4} }

\def\sumtaufoneetabavg{(1.26\pm 0.06\pm 0.06) \times 10^{-4} }
\def\sumtaufone{(3.60\pm 0.18\pm 0.23) \times 10^{-4} }

\def\sumetaexclusive{(0.99\pm 0.09\pm 0.13) \times 10^{-4} }

\def\ratiofourptoppe{0.28\pm 0.02\pm 0.02 }

\def\fitdfa{1.28031\pm 0.00067}
\def\fitdfb{1.27787\pm 0.00080}
\def\fitdfc{1.28036\pm 0.00335}
\def\fitdfd{1.27775\pm 0.00045}
\def\fitdcfa{1.28105\pm 0.00067}
\def\fitdcfb{1.27805\pm 0.00082}
\def\fitdcfc{1.28383\pm 0.00337}
\def\fitdcfd{1.28067\pm 0.00060}

\def\delmcfa{0.00074\pm 0.00008}
\def\delmcfb{0.00018\pm 0.00020}
\def\delmcfc{0.00347\pm 0.00033}
\def\delmcfd{0.00292\pm 0.00040}
\def\difmavg{-0.91\pm 0.10}
\def\mfr{1.28025\pm 0.00039}
\def\mf{1.28116\pm 0.00039\pm 0.00045}

\def\sigtt{\ensuremath{\sigma_{\tautau}}}

\begin{document}
%
\author{J.~P.~Lees}
\author{V.~Poireau}
\author{V.~Tisserand}
\affiliation{Laboratoire d'Annecy-le-Vieux de Physique des Particules (LAPP), Universit\'e de Savoie, CNRS/IN2P3,  F-74941 Annecy-Le-Vieux, France}
\author{J.~Garra~Tico}
\author{E.~Grauges}
\affiliation{Universitat de Barcelona, Facultat de Fisica, Departament ECM, E-08028 Barcelona, Spain }
\author{A.~Palano$^{ab}$ }
\affiliation{INFN Sezione di Bari$^{a}$; Dipartimento di Fisica, Universit\`a di Bari$^{b}$, I-70126 Bari, Italy }
\author{G.~Eigen}
\author{B.~Stugu}
\affiliation{University of Bergen, Institute of Physics, N-5007 Bergen, Norway }
\author{D.~N.~Brown}
\author{L.~T.~Kerth}
\author{Yu.~G.~Kolomensky}
\author{G.~Lynch}
\affiliation{Lawrence Berkeley National Laboratory and University of California, Berkeley, California 94720, USA }
% ...
\collaboration{The \babar\ Collaboration}
\noaffiliation

\title{
{\large \bf \boldmath   Study of high-multiplicity three-prong and five-prong  \mtau decays at \babar\ }
}
\begin{flushleft}
BaBar-PUB-12/008 \\
SLAC-PUB-15047 \\
\end{flushleft}


\vspace{1pc}
\begin{abstract}
\begin{center}
\large \bf Abstract
\end{center}
We present measurements of the branching fractions of three-prong and five-prong
\mtau decay modes using a sample of 430 million \mtau lepton pairs, corresponding
to an integrated luminosity of $468\,\invfb$, collected with the \babar\ 
detector at the \pep2\ asymmetric-energy \epem storage rings at SLAC. 
The \tauthreepieta, \tauthreepiomega, and 
$\taum \! \rightarrow \pim f_1(1285) \nut$ branching fractions
are presented, as well as a new limit on the branching fraction  
of the second-class current decay \tauetaprime.
We search for the decay mode \tauketaprime and for five-prong decay
modes with kaons, and place the first upper limits on their branching fractions.
\end{abstract}


\maketitle


\section{Introduction}
Study of the three-prong and five-prong decay modes of the \mtau lepton,
where ``prong'' refers to the number of charged hadrons ($\pi$ or \kaon) in
the final state, allows one to test the Standard Model and search for 
evidence of new physics.  
The large \mtau lepton data sample collected by the \babar\ experiment 
provides an opportunity to perform a comprehensive study of rare, high 
multiplicity decay modes and to search for forbidden processes.

We present measurements of the \tauthreepieta, \tauthreepiomega, and \taufpi
branching fractions. 
We use the primary decay modes of the $\eta$, $\omega(782)$, and $f_1(1258)$ mesons:
\etagg, \etappp, \etapiz, \omegappp, \fpppp, and \feta.
No other narrow resonances are observed. 
We measure the branching fractions of the non-resonant decays, where the non-resonant 
category includes possible contributions from broad resonances.
We present a new limit on the branching fractions of the second-class current decay
\tauetaprime, and the first limits on the allowed first-class 
current decays \tauketaprime and \taupietaprimepiz.
Finally, we present the first limits on the branching fractions of five-prong decay modes 
in which one or more of the charged hadrons is a charged kaon.
Note that the branching fractions exclude the contribution 
of $\KS \rightarrow \pip\pim$ decays.
Throughout this paper, charge conjugation is implied.

This analysis is based on data recorded 
with the \babar\ detector at the \pep2\ asymmetric-energy \epem\ 
storage rings operated at the SLAC National Accelerator Laboratory. 
With an integrated luminosity (\lum) of 424 and 44~\invfb recorded at 
center-of-mass (CM) energies of 10.58 \gev and 10.54 \gev, respectively,  
and an $\epem \rightarrow \taup \taum$ cross section of 
$\sigma_{\tautau} = (0.919\pm0.003)$ nb \cite{Banerjee:2007is}, the data sample
contains 430 million \mtau pairs.

The \babar\ detector is described in detail in Ref.~\cite{detector}.
Charged-particle  momenta are measured with a five-layer
double-sided silicon vertex tracker and a 40-layer drift chamber, 
both operating in the 1.5 T magnetic field of a superconducting solenoid.
Information from a detector of internally reflected Cerenkov light is 
used in conjunction with specific energy loss measurements from the 
tracking detectors to identify charged pions and kaons \cite{Aubert:2007mj}. 
Photons are reconstructed from energy clusters deposited
in a CsI(TI) electromagnetic calorimeter.  
Electrons are identified  by combining tracking and calorimeter information.
An instrumented magnetic flux return is used to identify muons.

The background contamination and selection efficiencies are determined
using Monte Carlo simulation.
The \mtau-pair production is simulated with the KK2F event
generator \cite{kk}. 
The \mtau decays, continuum \qqbar events ($q=udsc$), and final-state radiative 
effects are modeled with the Tauola \cite{tauola1}, JETSET \cite{qq1}, 
and Photos \cite{tauola2} generators, respectively. 
Dedicated samples of \tautau events are created using the 
Tauola or EVTGEN \cite{evtgen} 
programs, with one of the $\tau$ leptons allowed to decay to any mode while
the other $\tau$ decays to a specific final state. 
The detector response is simulated with GEANT4 \cite{GEANT}. 
All Monte Carlo events are processed through a full simulation of the \babar\ 
detector and are reconstructed in the same manner as the data.

\vspace{0.25cm}
\section{Event Selection}

The $\tau$ pair is produced back-to-back in the \epem CM frame.
The decay products of the two $\tau$ leptons can thus be separated from 
each other by dividing the event into two hemispheres: the ``signal'' 
hemisphere and the ``tag'' hemisphere. The separation is performed using the 
event thrust axis \cite{thrust},  
which is calculated using all charged particle and photon 
candidates in the event.

We select events where one hemisphere (tag) contains exactly one track 
while the 
other hemisphere (signal) contains exactly three 
or five tracks 
with total charge opposite to that of the tag hemisphere.
The event is rejected if any pair of oppositely charged tracks is
consistent with a photon conversion.
The component of the momentum transverse to the beam axis for 
each of the tracks must be greater than 0.1\gevc in the laboratory frame.  
All tracks are required to have a point of closest approach to the interaction region 
less than 1.5\cm in the plane transverse to the beam axis and 
less than 2.5\cm in the direction along that axis.  
This requirement eliminates \KS mesons that decay to $\pip\pim$ at points
distant from the \epem collision point.

To reduce backgrounds from non-\mtau-pair events, we require that the momentum 
of the charged particle in the tag hemisphere be less than 4\gevc in the 
CM frame and that the charged particle be identified as an electron or a muon.  
The \qqbar background is suppressed by requiring that
there be at most one energetic $(E > 1 \gev)$ electromagnetic calorimeter 
cluster in the tag hemisphere that is not associated with a track.
Additional background suppression is achieved by requiring 
the magnitude of the event thrust to lie between 0.92 and 0.99.

Neutral pion and eta candidates are reconstructed from two photon candidates, 
each with energy greater than 30~MeV in the laboratory frame; the 
invariant mass of the \piz ($\eta$) is required to be between 0.115 (0.35) and 
0.150 (0.70) \gevcc. 
Neutral pion candidates are reconstructed in the signal hemisphere.
If a photon candidate meets the invariant mass requirement with multiple 
photon candidates, then the neutral pion candidate with invariant mass 
closest to the nominal \piz mass \cite{pdg} is selected. 
The search for additional pion candidates is repeated using the remaining
photon candidates.
The residual photon clusters in the signal hemisphere are used to search 
for \etagg candidates. 
In the case of multiple $\eta$ candidates, the candidate with invariant 
mass closest to the nominal $\eta$ mass is selected. 
We reject events in which the invariant mass $M$ formed from the system of 
charged particles, \piz, and $\eta$ candidates, all in the signal hemisphere, 
exceeds 1.8\gevcc.

The branching fractions  are calculated using the expression 
$\mathcal{B} = N_{X} / (2 N_{\tt} \varepsilon)$
where $N_{X}$ is the number of candidates after background subtraction, 
$N_{\tt}$ is the number of \mtau pairs produced,
and $\varepsilon$ is the selection efficiency.
$N_{\tt}$ is determined from the product of the integrated 
luminosity and the $\epem \rightarrow \tautau$ cross section. 
The uncertainty of $N_{\tt}$ is estimated to be 1\%.
The selection efficiencies are determined from the signal 
Monte Carlo samples.
The uncertainty on the selection efficiencies includes 0.5\% per track 
on the track reconstruction efficiency, as well as particle identification (PID) 
selection uncertainties.
From studies conducted on real and simulated events, the uncertainties on the charged 
particle identification selectors are estimated
to be 1\% for electrons, 2.5\% for muons,  0.5\% for pions, and 1.8\% for kaons.
The combined electron and muon particle identification uncertainty is estimated
to be 1.6\% based on the composition of the event samples.
The uncertainty on the $\piz \rightarrow \gamma\gamma$ and 
\etagg reconstruction efficiency is estimated to be 3\% per candidate.

\begin{figure}[htpb]
\begin{center}
\mbox{\epsfig{file=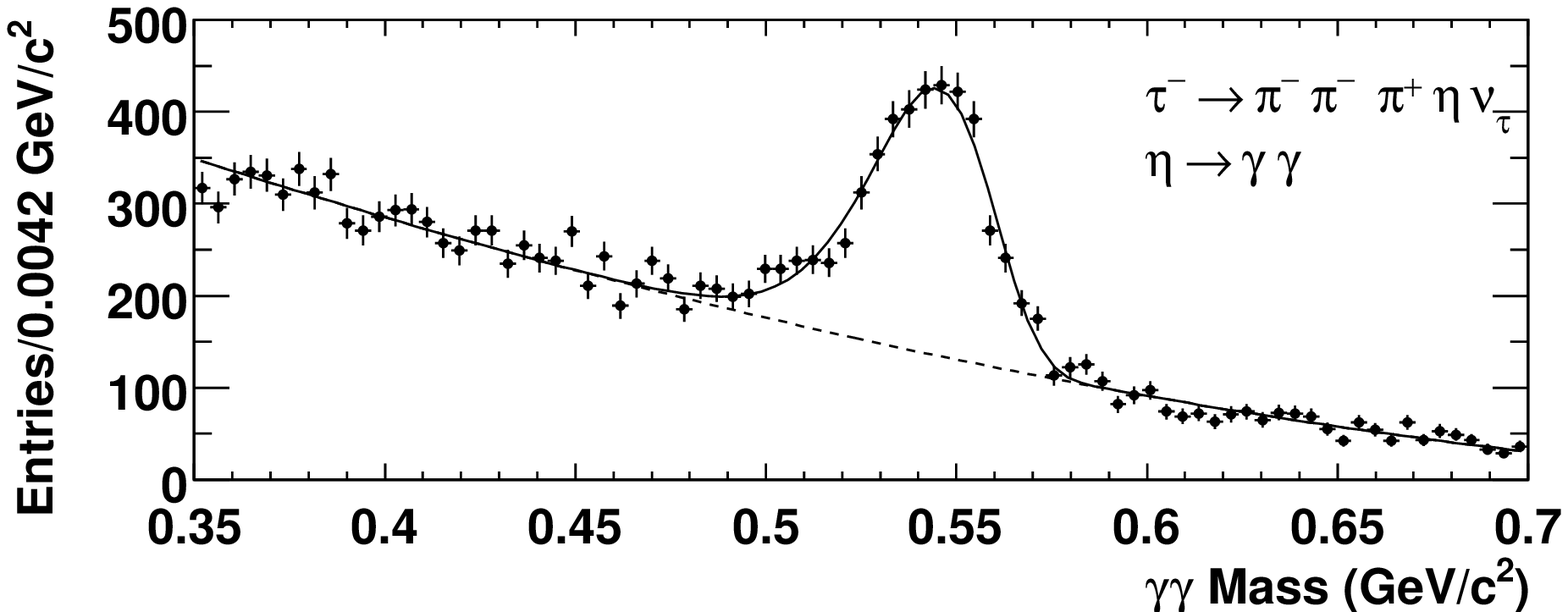,          height=3.2cm}} 
\mbox{\epsfig{file=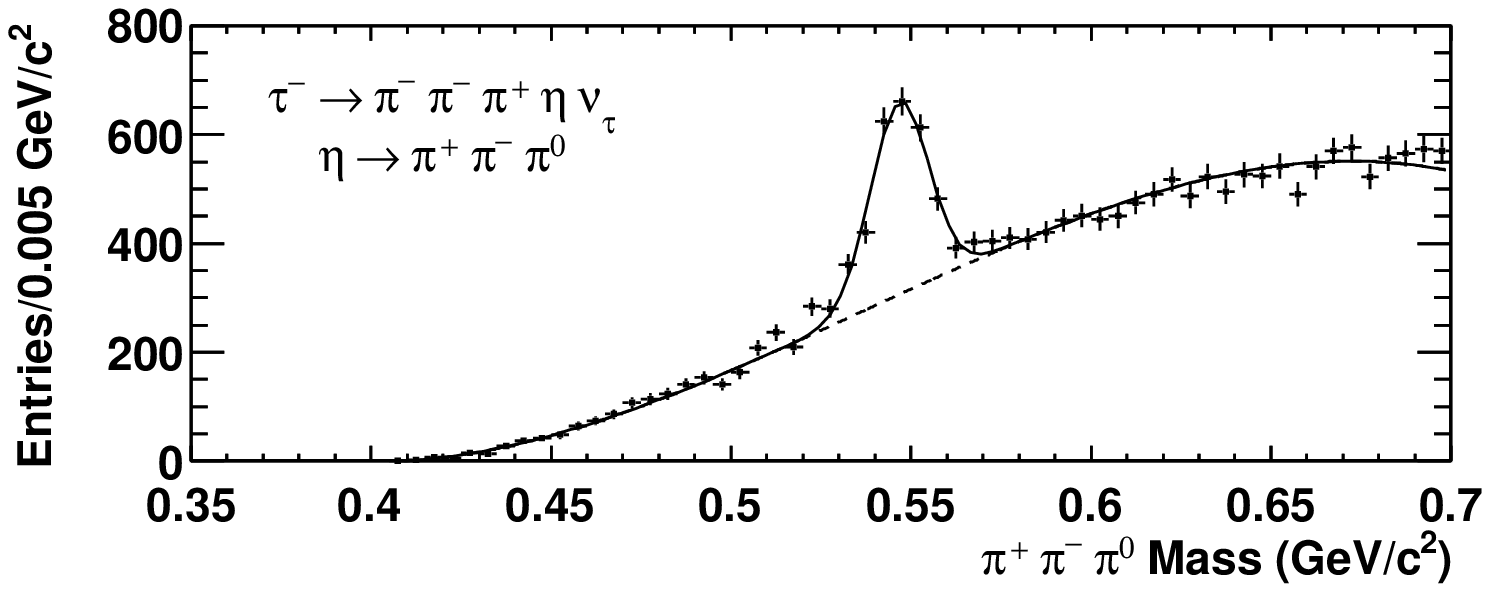,      height=3.2cm}} 
\mbox{\epsfig{file=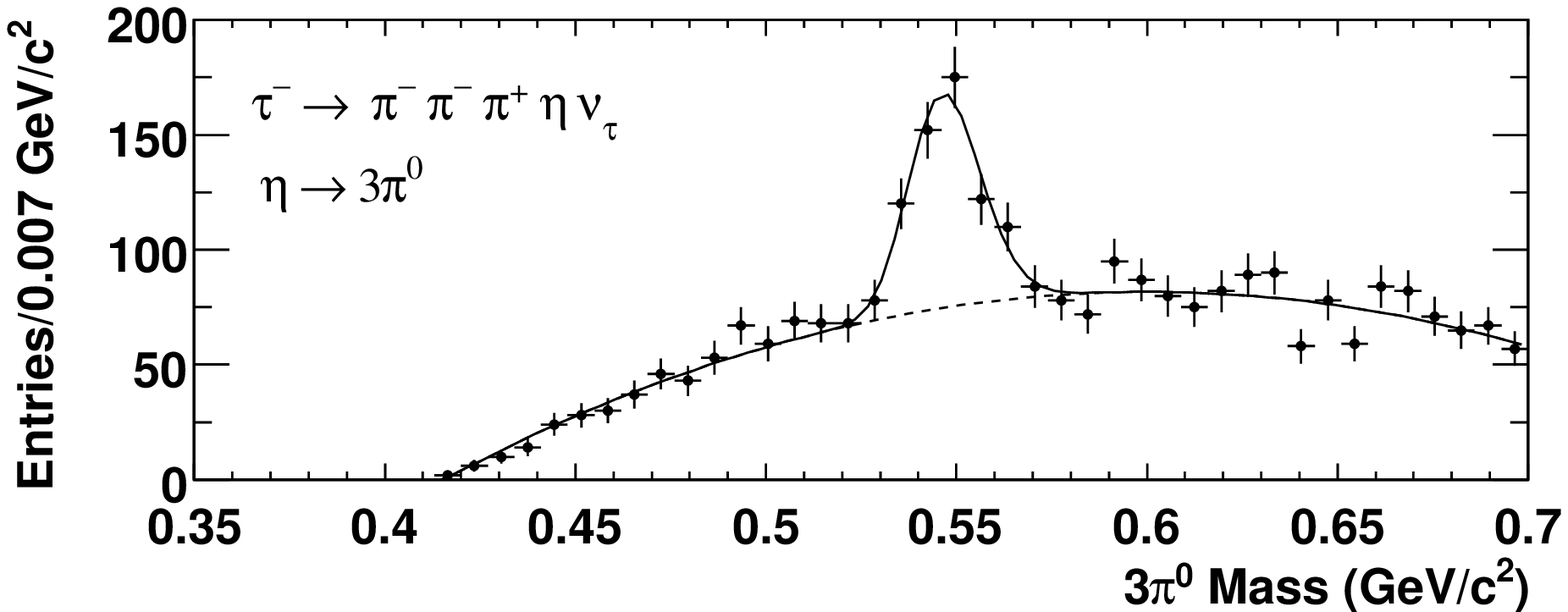,        height=3.2cm}} 
\mbox{\epsfig{file=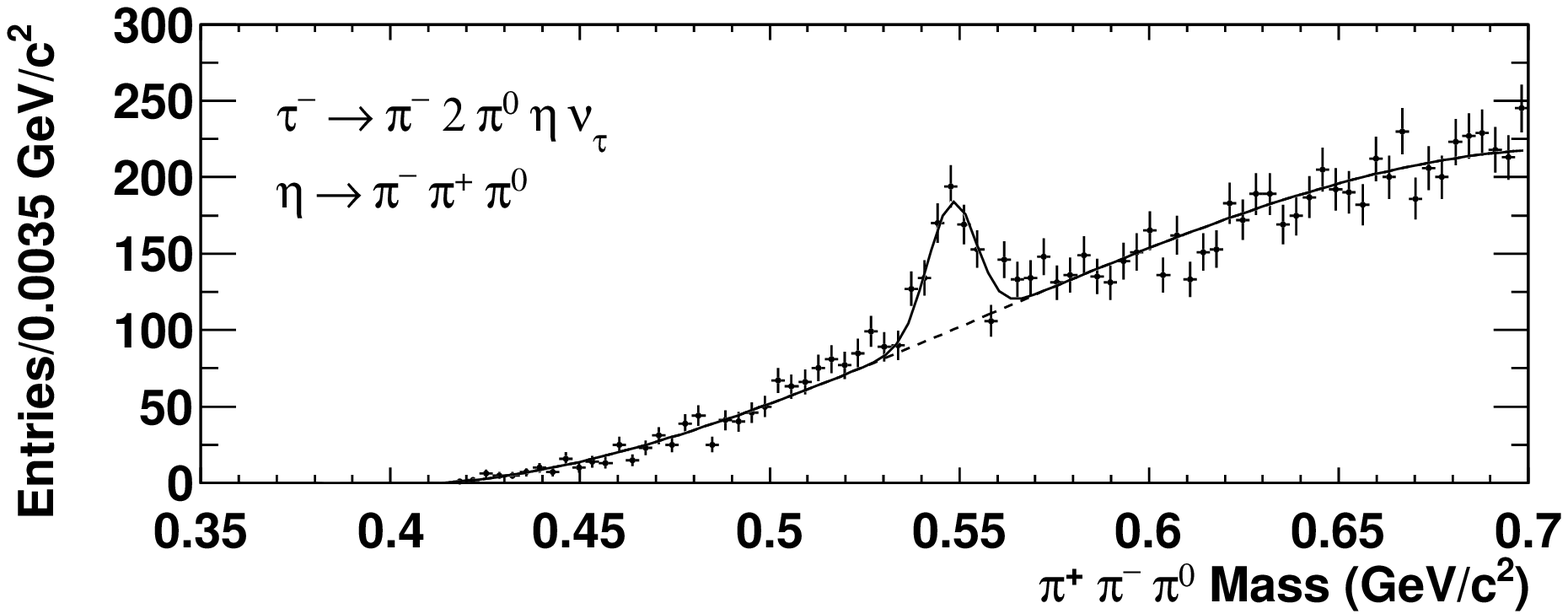,  height=3.2cm}}
\end{center}
\caption{\label{fig:eta}
The  $\gamma\gamma$, $\pip\pim\piz$,  and 3\piz invariant mass distributions 
for \taupppe decay candidates, and the  $\pip\pim\piz$  invariant mass distribution for
\taupzzeta decay candidates,  after all selection criteria are applied.
The solid lines represent the fit to the $\eta$ peak and background.
The dashed lines show the extrapolation of the background function under the
$\eta$ peak.
}
\end{figure}

\vspace{0.25cm}
\section{Results}

We present measurements of \mtau decays to a system  with $\eta$, $f_1$ and 
$\omega$ resonances in Sections  A, B,  and C, respectively.  
Decays with these resonances do not account for all three-prong or five-prong 
\mtau decay modes, as discussed below, and we present measurements of 
the \mtau  branching fractions through non-resonant modes in Section D.
Section E presents a search for $\tau$ decays with an \etapr(958) meson, 
while Section F presents a search for decays with either one or two charged kaons.

\begin{table*}[htpb]
\renewcommand\arraystretch{1.25}
\begin{center}
\caption{\label{table:eta} 
Results and branching fractions for \tauthreepieta decays.}
\vspace{0.25cm}
\begin{tabular}{lcccc} \hline \hline
 Decay Mode          & \taupppe & \taupppe  & \taupppe  & \taupzzeta \\
                     & \etagg   & \etappp   & \etapiz   & \etappp    \\ 
\hline 
Branching fraction ($10^{-4}$)
           & \hspace{0.2cm} $\tauetaBRGamCorrNoBra$      \hspace{0.2cm}
           & \hspace{0.2cm} $\etabrx$                    \hspace{0.2cm}
           & \hspace{0.2cm} $\BRThreePizNoBra$           \hspace{0.2cm} 
           & \hspace{0.2cm} $\tauetaCorrBRPiTwoPizNoBra$ \hspace{0.2cm} \\
Data events          & $\NsigEtaGam$           
                     & $\etanevt$ 
                     & $\NsigEtaPiz$ 
                     & $\NsigEtaPiTwoPiz$  \\ 
$\chi^2/NDF$         & $\EtaDataChiGam$        
                     & $\etachisq/\etandf$ 
                     & $\EtaDataChiPiz$ 
                     & $\EtaDataChiPiTwoPiz$ \\
Selection efficiency & $\EtaEffGam$            
                     & $\etaeffp$\%  
                     & $\EtaEffPiz$ 
                     & $\EtaEffPiTwoPiz$ \\
Background events    & $\NqqEtaMCGam$          
                     & $\etanqq$  
                     & $\NqqEtaMCPiz$ 
                     & $\NbkgEtaPiTwoPiz$ \\
\hline 
Systematic uncertainties (\%)    & & & \\
Tracking efficiency  & \EtaGammaTrackingSyst 
                     & \etasa 
                     & \EtaThreePiZeroTrackingSyst 
                     & \EtaPiTwoPiZeroTrackingSyst \\
\piz and $\eta$ PID  & \EtaGammaEtaEffSyst   
                     & \etasb 
                     & \EtaThreePiZeroEtaEffSyst 
                     & \EtaPiTwoPiZeroEtaEffSyst \\
Pion PID             & \EtaGammaPiPIDSyst    
                     & \etasc 
                     & \EtaThreePiZeroPiPIDSyst
                     & \EtaPiTwoPiZeroPiPIDSyst \\
Lepton-tag PID       & \EtaGammaPIDSyst      
                     & \etasd 
                     & \EtaThreePiZeroPIDSyst 
                     & \EtaPiTwoPiZeroSystErrLepPID \\
$N_{\tt}$            & \EtaGammaLumiSyst     
                     & \etasf 
                     & \EtaThreePiZeroLumiSyst  
                     & \EtaPiTwoPiZeroLumiSyst \\
Selection efficiency & 3.9    
                     & \etash 
                     & \EtaThreePiZeroMCEffSyst  
                     & \EtaPiTwoPiZeroMCEffSyst \\
Background           & \EtaGammaEtaQQSyst
                     & \etasg 
                     & \EtaThreePiZeroBkgSyst 
                     & \EtaPiTwoPiZeroBkgSyst \\
$\mathcal{B}$(\etagg)      & \EtaGammaBgammaSyst & - & - & - \\
$\mathcal{B}$(\etappp)     & - & \etase & - &\EtaPiTwoPiZeroBetapipipizSyst \\
$\mathcal{B}$(\etapiz)     & - & - & \EtaThreePiZeroBpizSyst & - \\
\hline 
Total (\%)           & \EtaGammaSystTotal    
                     & \etass 
                     & \EtaThreePiZeroSystTotal 
                     & \EtaPiTwoPiZeroSystTotal \\
\hline\hline
\end{tabular}
\end{center}
\end{table*}

\subsection{\tauthreepieta}

The \taupppe mode is studied in the \etagg, \etappp, and \etapiz final states, while
the \taupzzeta mode is studied in the \etappp final state.

The event yields are determined by fitting the $\eta$ mass peak in the 
$\gamma\gamma$, $\pip\pim\piz$, and 3\piz invariant mass distributions  
(see Fig.~\ref{fig:eta}).
The fit uses a Novosibirsk function \cite{novo}
(a Gaussian distribution with a tail parameter) for the $\eta$ and a 
polynomial function for the background.

The Monte Carlo simulation indicates that some of the entries in the 
$\eta$ peak are from $\epem \rightarrow \qqbar$ events.
Control samples, obtained by reversing the requirement on the invariant mass 
of the observed decay products ($M > 1.8 \gevcc$), are used to validate the 
background estimate.
The expected background is corrected by the ratio of data to Monte Carlo events, 
and the statistical uncertainty of the ratio is included in 
the background systematic uncertainty.
This method of validating the \qqbar background estimate is used for all decays
and is not mentioned in the later sections.

The reconstruction efficiencies are determined from fits to the signal 
Monte Carlo samples.
The \taupzzeta sample is generated using a phase-space model for the 
final-state particles.
The \taupppe sample is composed of \taufpi (\feta) decays and decays
without an intermediate resonance.
The \taupppe (excluding $f_1$) and \taufpi efficiencies are the
same for  \etappp and \etapiz events, whereas a slight difference is observed 
for \etagg events and a 2.5\% uncertainty is added to the selection efficiency 
systematic for this mode.
In addition, a 4\% uncertainty is added to the \taupppe selection 
efficiency for the \etappp mode to take into account variations
observed for different fits.

The three determinations of the \taupppe branching fraction are found to be in good 
agreement (see Table~\ref{table:eta}) and we therefore calculate a 
weighted average. The statistical and systematic uncertainties 
on the average are obtained by combining the individual uncertainties 
in quadrature, accounting for correlations between the systematic terms. 
The weighted average (inclusive of \taufpi) is found to be
\begin{eqnarray*}
\mathcal{B}(\taupppe) & = \sumetabavg .
\end{eqnarray*}
Hereinafter, when two uncertainties are quoted, the first is statistical and 
the second is systematic.
The average branching fraction (exclusive of \taufpi) is 
determined to be $\sumetaexclusive$
and is obtained using the branching fraction (inclusive of \taufpi),
given above, and subtracting the product branching fraction
$\mathcal{B}(\taufpi) \times \mathcal{B}(\feta)$ presented in the next section.

The \taupzzeta branching fraction is found to be
\begin{eqnarray*}
\mathcal{B}(\taupzzeta) & = \tauetaCorrBRPiTwoPiz .
\end{eqnarray*}
Naively, the ratio of the \taupppe to \taupzzeta branching 
fractions is expected to be two if the decay is dominated by the \taufpi 
decay mode (based on the $f_1$ branching fractions \cite{pdg}).
The data do not support this expectation.

Our previous measurement of the \taupppe branching fraction
$(1.60 \pm 0.05 \pm 0.11) \times 10^{-4}$ \cite{babar:3pieta}, which is based 
on the \etagg mode only, is superseded by this measurement. 
The fit used in the previous analysis was performed using
a narrower range in the invariant mass distribution, and a background
parametrization that is less accurate than in the present study.

The \taupppe and \taupzzeta branching fractions are in good agreement
with the results from the CLEO Collaboration, 
$(2.3 \pm 0.5) \times 10^{-4}$ 
and  
$(1.5 \pm 0.5) \times 10^{-4}$,
respectively \cite{cleo:sixpi}.
Li predicts a larger \taupppe branching fraction, 
$2.93 \times 10^{-4}$  \cite{theory:li}.
 
\begin{figure}
\begin{center}
\mbox{\epsfig{file=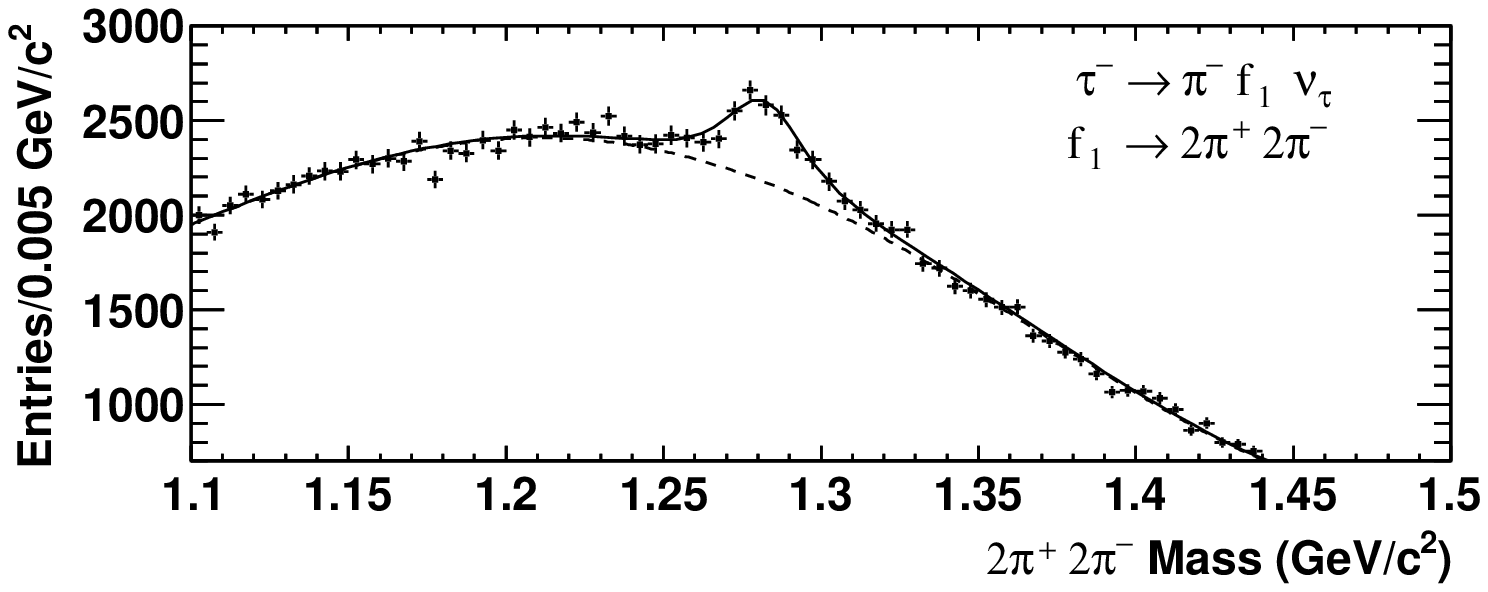,height=3.2cm}}
\mbox{\epsfig{file=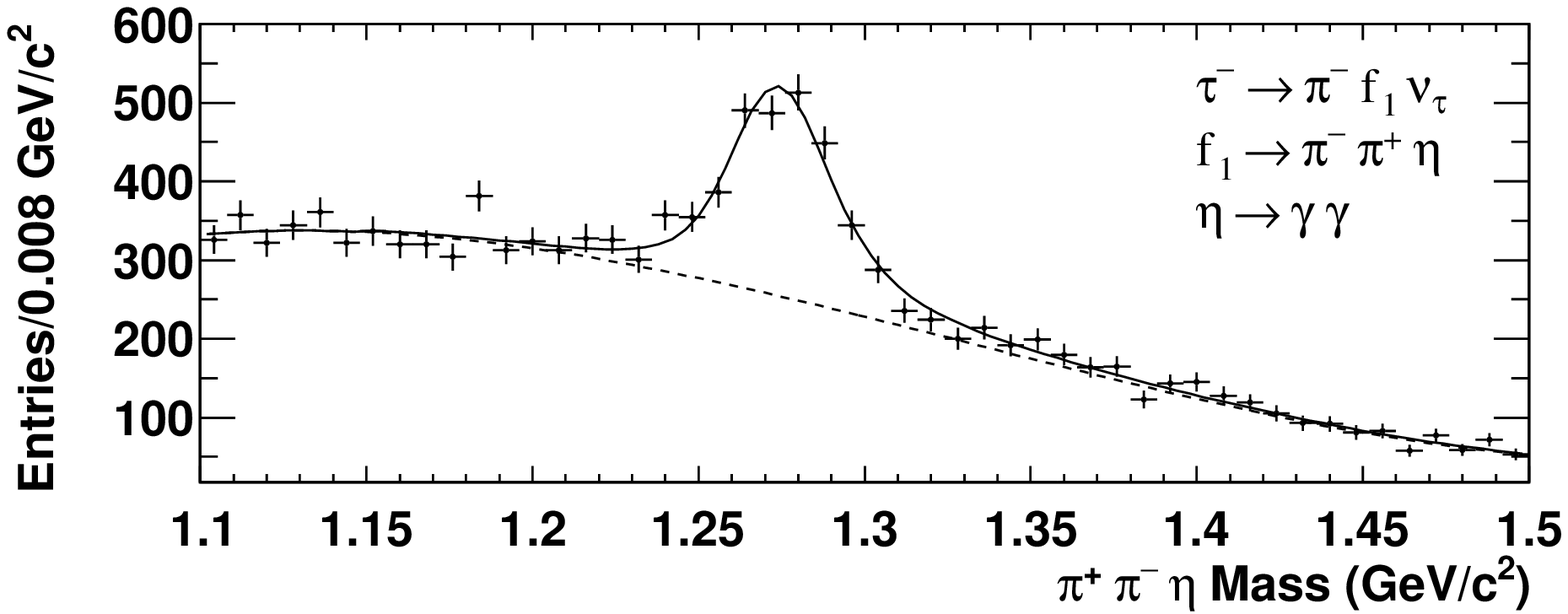,    height=3.2cm}} 
\mbox{\epsfig{file=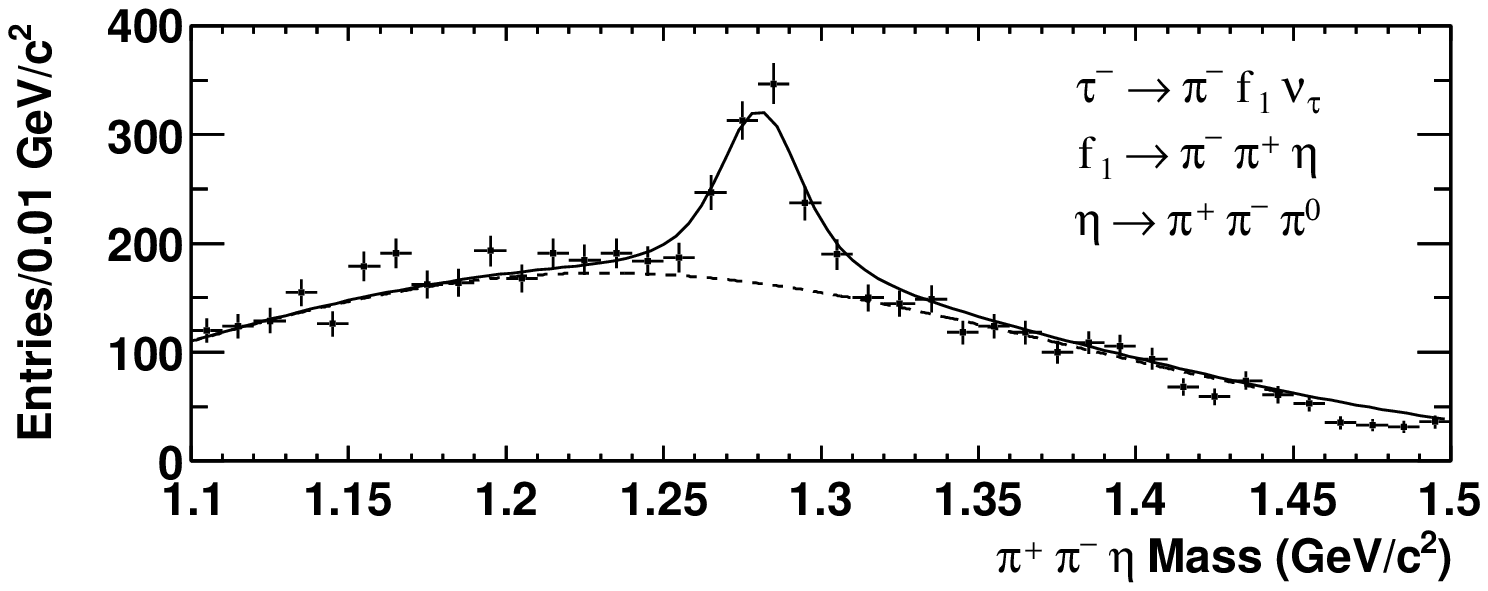,height=3.2cm}} 
\mbox{\epsfig{file=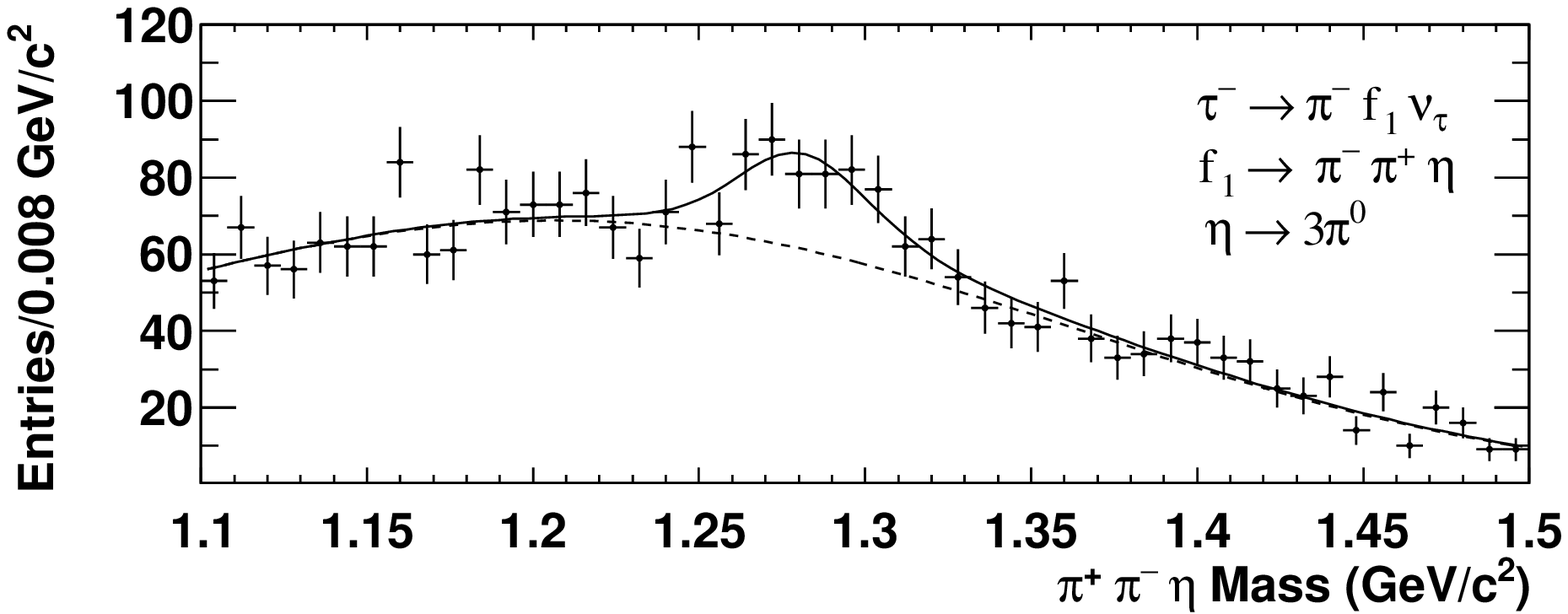,  height=3.2cm}}  
\end{center}
\caption{\label{fig:f1}
The 2\pip 2\pim and $\pip \pim \eta$ invariant mass distributions 
for \taupppe decay candidates after all selection criteria are applied.
The lower three plots are for the \etagg, \etappp, and \etapiz decays.
The solid lines represent the fit to the $f_1(1285)$ peak and background. 
The dashed lines show the extrapolation of the background function under the
$f_1$ peak.
}
\end{figure}

\begin{figure}[htpb]
\begin{center}
\mbox{\epsfig{file=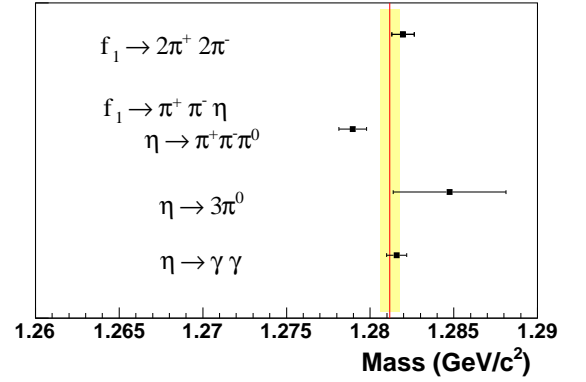,height=6cm}}
\end{center}
\caption{\label{fig:fonemassfit}
Compilation of our measurements of the $f_1$ mass.
The solid line is the weighted average and the shaded area is the one-standard-deviation
region.
}
\end{figure}

\begin{figure}[htpb]
\begin{center}
\mbox{\epsfig{file=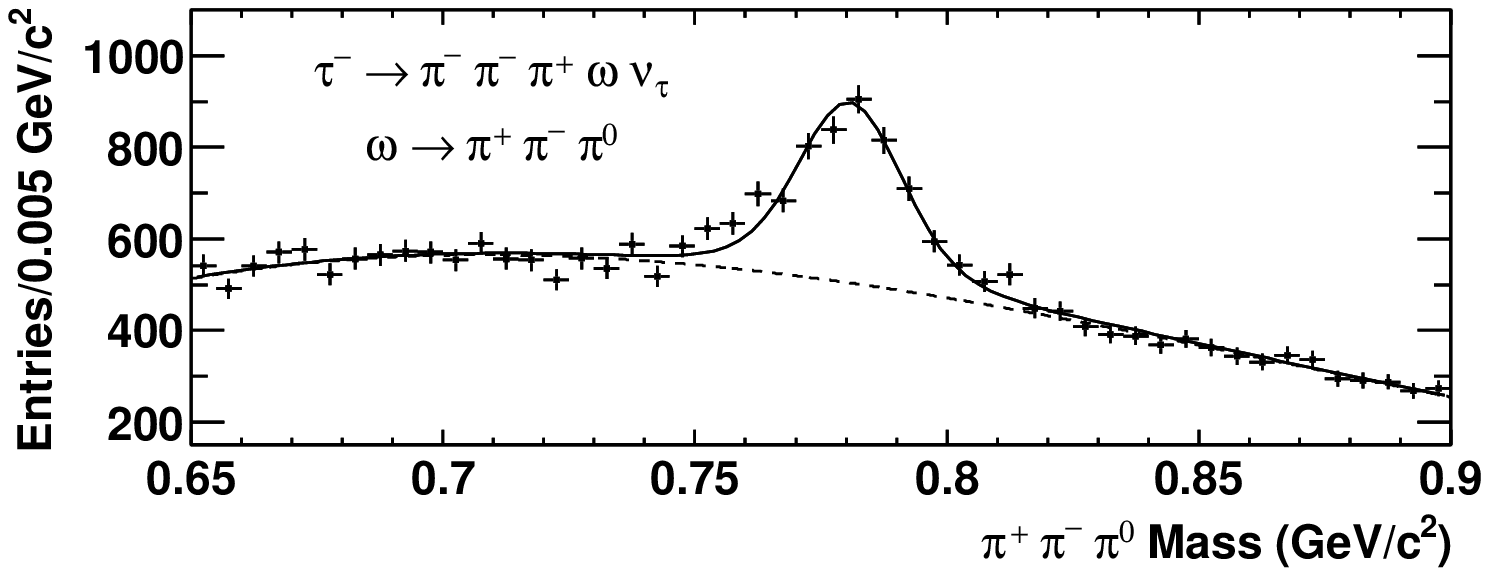,    height=3.4cm}}
\mbox{\epsfig{file=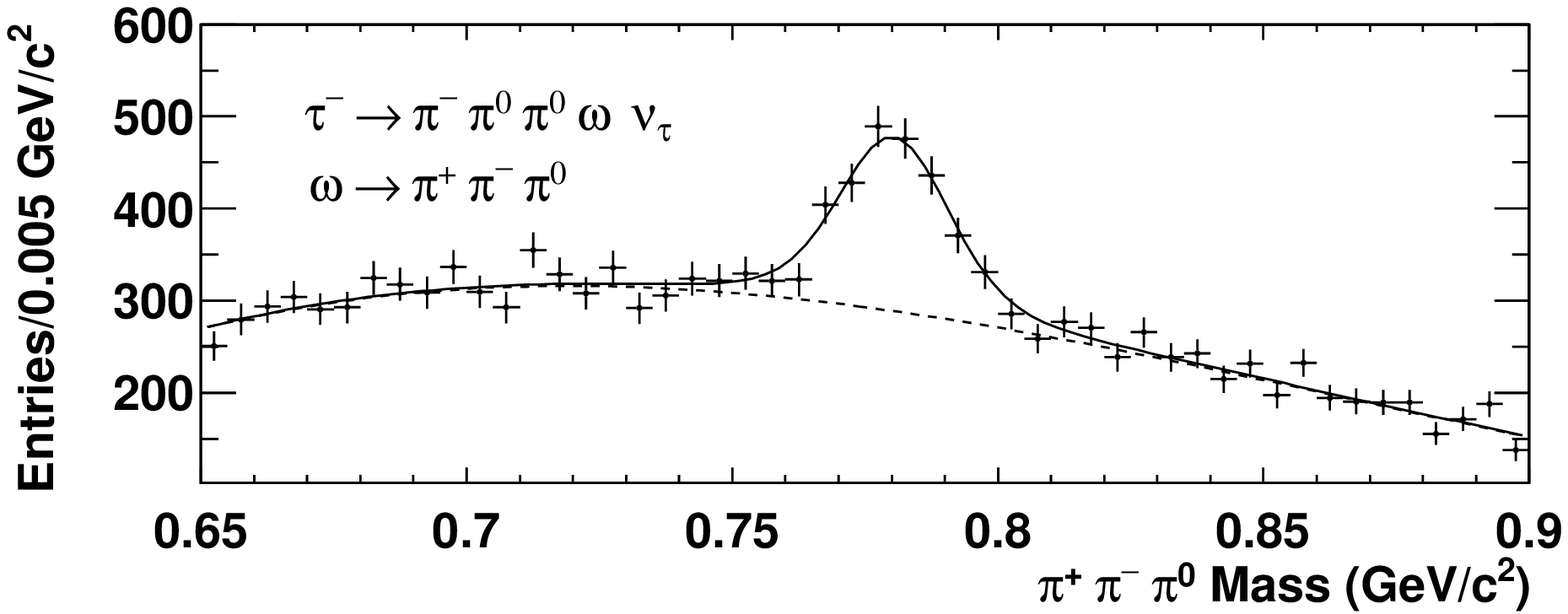,height=3.4cm}}
\end{center}
\caption{\label{fig:omega}
The fits to the $\omega$ peak in the $\pip\pim\piz$
invariant mass distributions for \taupppo and \taupzzomega decay candidates 
after all selection criteria are applied.
The solid lines represent the fit to the $\omega$ peak and background. 
The dashed lines show the extrapolation of the background function under the
$\omega$ peak.
}
\end{figure}

\begin{table*}[htpb]
\renewcommand\arraystretch{1.25}
\begin{center}
\caption{\label{table:f1} 
Results and branching fractions for \taufpi decays.}
\vspace{0.25cm}
\begin{tabular}{lcccc} \hline \hline
Decay Mode           & \fpppp   & \feta  & \feta   & \feta    \\ 
                     &          & \etagg & \etappp & \etapiz  \\ 
\hline 
Branching fractions ($10^{-4}$)  \hspace{0.5cm} & & & & \\
$\mathcal{B}(\taufpi) \mathcal{B}(f_1 \rightarrow 2\pip 2\pim)$ & $\ffbrexx$ & & & \\
$\mathcal{B}(\taufpi) \mathcal{B}(f_1\rightarrow \pi^-\pi^+\eta)$      & 
         & \hspace{0.25cm} $\taufBRnofetaGamNoBra$  \hspace{0.25cm}
         & \hspace{0.25cm} $\fonebrbx$            \hspace{0.25cm}
         & \hspace{0.25cm} $\taufBRnofetaPizNoBra$  \hspace{0.25cm} \\
\hline
Data events          & $\ffadnevt$            
                     & $\NsigfGam$            
                     & $\fonenevt$ 
                     & $\NsigfPiz$ \\ 
$\chi^2/NDF$         & $\ffadchisq/\ffadndf$ 
                     & $\fDataChiGam$        
                     & $\fonechisq /\fonendf$ 
                     & $\fDataChiPiz$ \\
Selection efficiency & $\ffeffp$               
                     & $\fEffGam$            
                     & $\fgeffp$  
                     & $\fEffPiz$ \\ 
\hline 
Systematic uncertainties (\%)    & & & & \\
Tracking efficiency  & \ffssca 
                     & \FoneEtaGammaTrackingSyst  
                     & \fgssa 
                     & \FoneEtaThreePizTrackingSyst \\
\piz and $\eta$ PID  & - 
                     & \FoneEtaGammaEtaEffSyst   
                     & \fgssb 
                     & \FoneEtaThreePizEtaEffSyst \\
Pion PID             & \ffsscb 
                     & \FoneEtaGammaPiPIDSyst    
                     & \fgssd 
                     & \FoneEtaThreePizPiPIDSyst \\
Lepton-tag PID       & \ffsscc 
                     & \FoneEtaGammaPIDSyst      
                     & \fgssc 
                     & \FoneEtaThreePizPIDSyst \\
$N_{\tt}$            & \ffsscd 
                     & \FoneEtaGammaLumiSyst     
                     & \fgsse 
                     & \FoneEtaThreePizLumiSyst\\ 
Selection efficiency & \ffsuc  
                     & \FoneEtaGammaMCEffSyst    
                     & \fgssg      
                     & \FoneEtaThreePizMCEffSyst \\
Fit model
                     & \ffsua  &  \FoneEtaGammaModeling & - & -\\ 
$\mathcal{B}$(\etagg)      & - & \FoneEtaGammaBgammaSyst & - & -\\
$\mathcal{B}$(\etappp)     & - & - &\fgssf & -\\
$\mathcal{B}$(\etapiz)     & - & - & - & \FoneEtaThreePizBpizSyst \\
\hline 
Total  (\%)          & \ffssex 
                     & \FoneEtaGammaSystTotal     
                     & \fgss 
                     & \FoneEtaThreePizSystTotal \\ 
\hline \hline
\end{tabular}
\end{center}
\end{table*}

\begin{table*}[htpb]
\renewcommand\arraystretch{1.25}
\begin{center}
\caption{\label{table:f1mass}
Results of fits for the mass of the $f_1$ resonance in \taufpi decays. 
The errors are statistical.}
\vspace{0.25cm}
\begin{tabular}{lccc}  
\hline \hline
Decay Mode               & Monte Carlo   
                         & Data 
                         & Data \\
                         & (generator - fit)
                         & (fit) 
                         & (corrected) \\
                         & (\gevcc)   
                         & (\gevcc)   
                         & (\gevcc)  \\
\hline 
\fpppp
                         & \hspace{0.25cm} $\delmcfa$ \hspace{0.25cm}
                         & \hspace{0.25cm} $\fitdfa$  \hspace{0.25cm}
                         & \hspace{0.25cm} $\fitdcfa$ \hspace{0.25cm} \\
\multicolumn{4}{l}{\feta}  \\ 
\hspace{0.5cm} 
\etagg
                         & $\delmcfd$ 
                         & $\fitdfd$ 
                         & $\fitdcfd$ \\
\hspace{0.5cm} 
\etappp
                         & $\delmcfb$ 
                         & $\fitdfb$ 
                         & $\fitdcfb$ \\
\hspace{0.5cm} 
\etapiz
                         & $\delmcfc$ 
                         & $\fitdfc$ 
                         & $\fitdcfc$ \\
\hline \hline
\end{tabular}
\end{center}
\end{table*}

\subsection{\taufpi}

The branching fraction of \taufpi and the mass of the $f_1$ meson
are measured using the \fpppp and \feta decay modes, where the \feta decay 
is reconstructed using \etagg, \etappp, and \etapiz events.
The criteria used to select the \taufpi decays for the branching fraction
measurement are described earlier.
We modify the selection for the mass measurement, dropping  
the requirement that the track in the tag hemisphere be a lepton
and the restriction on the number of photon candidates in the tag hemisphere,
to increase the size of the event sample.

The numbers of \taufpi candidates are determined by fitting the $f_1$ peak in 
the $2\pip2\pim$ and $\pip\pim\eta$ invariant mass distributions 
(see Fig.~\ref{fig:f1}).
The $f_1$ lineshape is expected to be a Breit-Wigner distribution,
modified by the limited phase space.
Previous studies show that the \fak (\azeta) channel appears to account for all
\feta decays \cite{L3paper}.
The mass of the $\pi a_0 (980)$ system and the \mtau mass provide a lower 
and upper limit, respectively, on the  $f_1$ lineshape.
We use  the four-vectors of  the charged pion and $a_0 (980)$ from the EVTGEN
generator to determine the simulated $f_1$ lineshape and find it to 
be a close approximation to the Breit-Wigner expectation.
The $f_1$ peak is fit using this lineshape convolved with a Gaussian distribution
to take into account the effects of the detector resolution. 
The results of the fits are presented in Table~\ref{table:f1}.
There is no evidence for peaking background from \qqbar events 
or other \mtau decays. 

The product of the \taufpi and \fpppp branching fractions, 
and the product of the \taufpi and \feta branching fractions, 
are measured to be
\begin{eqnarray*}
\mathcal{B}(\taufpi) \mathcal{B}(\fpppp) \\
= \ffbrex, \\
\mathcal{B}(\taufpi) \mathcal{B}(\feta) \\
= \sumtaufoneetabavg,
\end{eqnarray*}
respectively, where the second result is the weighted average 
of the three $\eta$ modes.
The $\mathcal{B}(\taufpi)$ branching fraction is determined to be
$\ffbr$ and $\sumtaufone$, as obtained by dividing
the product branching fractions by  
$\mathcal{B}$(\fpppp) = $0.110^{+0.007}_{-0.006}$ and 
$\mathcal{B}(\feta) = 0.349^{+0.013}_{-0.015}$
\cite{pdg:fone}, respectively.

Our two measured values for the \taufpi branching fraction are consistent with each other
to within two standard deviations of the combined statistical
and systematic uncertainties.
The ratio of the product branching fractions is used to determine
the ratio of the \fpppp and \feta branching fractions as
\begin{eqnarray*}
\frac{ \mathcal{B}(\fpppp) } {\mathcal{B}(f_1 \rightarrow \pi \pi \eta) } = 
\ratiofourptoppe, 
\end{eqnarray*}
where $ \mathcal{B}(f_1 \rightarrow \pi \pi \eta) = 1.5 \times \mathcal{B}(\feta)$
based on isospin symmetry.
This agrees with average value of $0.41 \pm 0.14$ quoted by the Particle Data
Group \cite{pdg} but disagrees with their fit value of $0.63 \pm 0.06$ \cite{pdg}.

The systematic uncertainties of the branching fractions are listed in Table~\ref{table:f1}.
We observe that the number of events in the $f_1$ peak in the \fpppp sample 
varies by 5\% for different background shapes. 
This variation is included as a systematic uncertainty.
We also observe that the selection efficiency obtained from the Monte Carlo simulation
exhibits a slight dependence on whether the $f_1$ decays via the \fak or the \feta
mode, and the variation is included as a systematic uncertainty (listed under 
``Fit model'' in Table~\ref{table:f1}).

The \taufpi branching fraction using the \fpppp mode is consistent with the 
previous \babar\ measurement, which is also based on the \fpppp mode
\cite{babar:5prong}.
CLEO published a branching fraction of
$(5.8^{+1.4}_{-1.3}\pm 1.8) \times 10^{-4}$ \cite{cleo:3pieta} and
Li predicts a branching fraction of  $2.9 \times 10^{-4}$
\cite{theory:li:fone}.

The $f_1$ mass is determined by fitting the peak with
a non-relativistic Breit-Wigner function, which was used in previous
measurements of the $f_1$ mass \cite{pdg}.
As a cross check, we use the energy-momentum four-vectors from the generator 
Monte Carlo simulation, and we find the fitted mass value to be consistent with 
the input mass value. 

We fit the invariant mass distribution in the fully reconstructed Monte Carlo 
samples to determine whether the result of
the fit differs from the input mass of the Monte Carlo 
generator. 
The difference is used to correct the value of the invariant mass of each channel
obtained from the fit and the uncertainty in the difference is included as a
systematic error.

Table~\ref{table:f1mass} and Fig.~\ref{fig:fonemassfit} show the 
results of the fits to the data.
The last column of the table presents the mass after the application of
the reconstruction correction factor.
The average of these results is $M_{f_1} = (\mfr) \gevcc$, where the error is statistical.

Previous \babar\ analyses have measured the invariant mass of resonances 
to be approximately 1\mevcc less than the values of Ref.~\cite{pdg}.
This shift is observed in the measurement of the mass of the 
$f_1$ meson \cite{babar:isromega} and the $\tau$ lepton \cite{babar:tau-mass}.
The shift is attributed to the absolute energy and momentum calibration of 
the detector.
We measure the calibration correction factor by fitting the $\eta$, $\omega$, 
\etapr, \Dz and \Dstarm states using data samples that have one track
in the tag hemisphere and either three or five tracks in the signal hemisphere.
No other selection criteria are applied.
The peak masses are found to be lower than the known values by 
$(\difmavg)$ \mevcc and the values are independent of the mass of the resonance.
The calibration correction factor is applied to the invariant mass 
and its error is included in the systematic uncertainty.

We determine the mass of the $f_1(1258)$ meson to be 
\begin{eqnarray*}
M_{f_1}  & = (\mf) \gevcc. 
\end{eqnarray*}
The systematic uncertainty includes the reconstruction uncertainty
and the calibration uncertainty.
This result is in good agreement with the value
$(1.2818 \pm 0.0006)$ \gevcc in Ref.~\cite{pdg}.

\begin{figure}[htpb]
\centering
\mbox{\epsfig{file=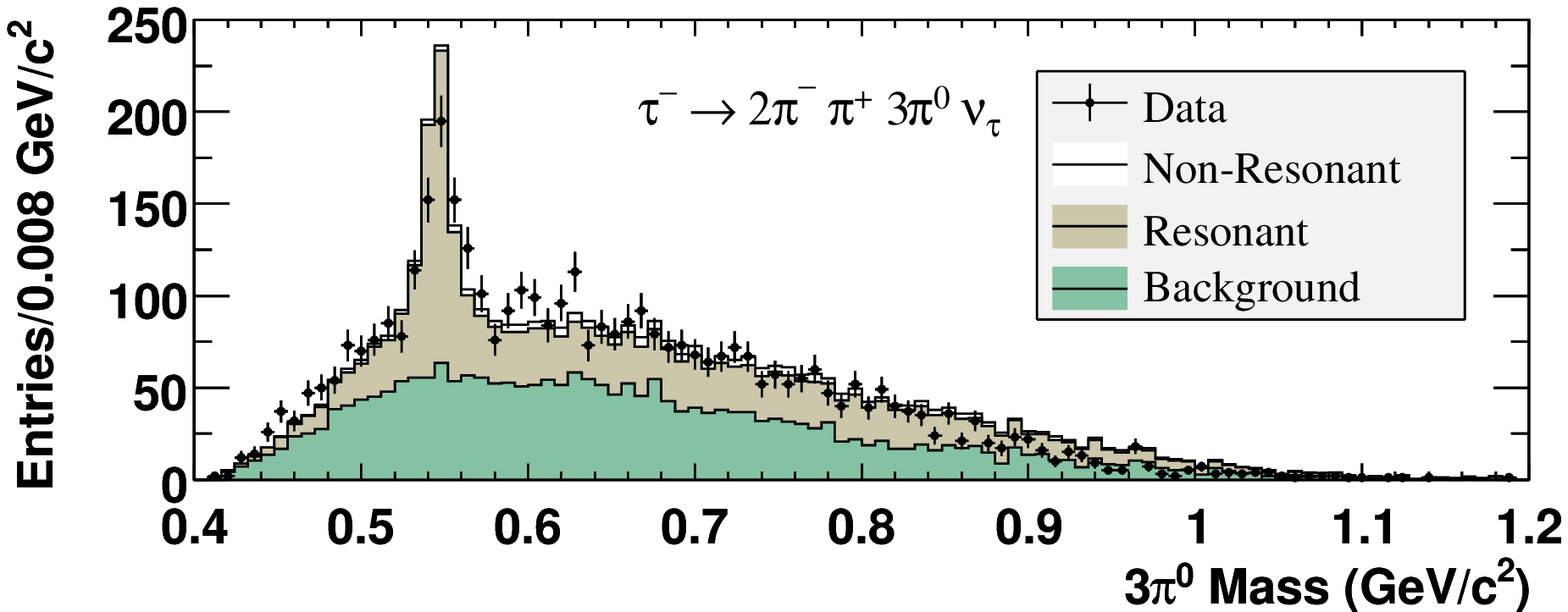,   height=3.4cm}}
\mbox{\epsfig{file=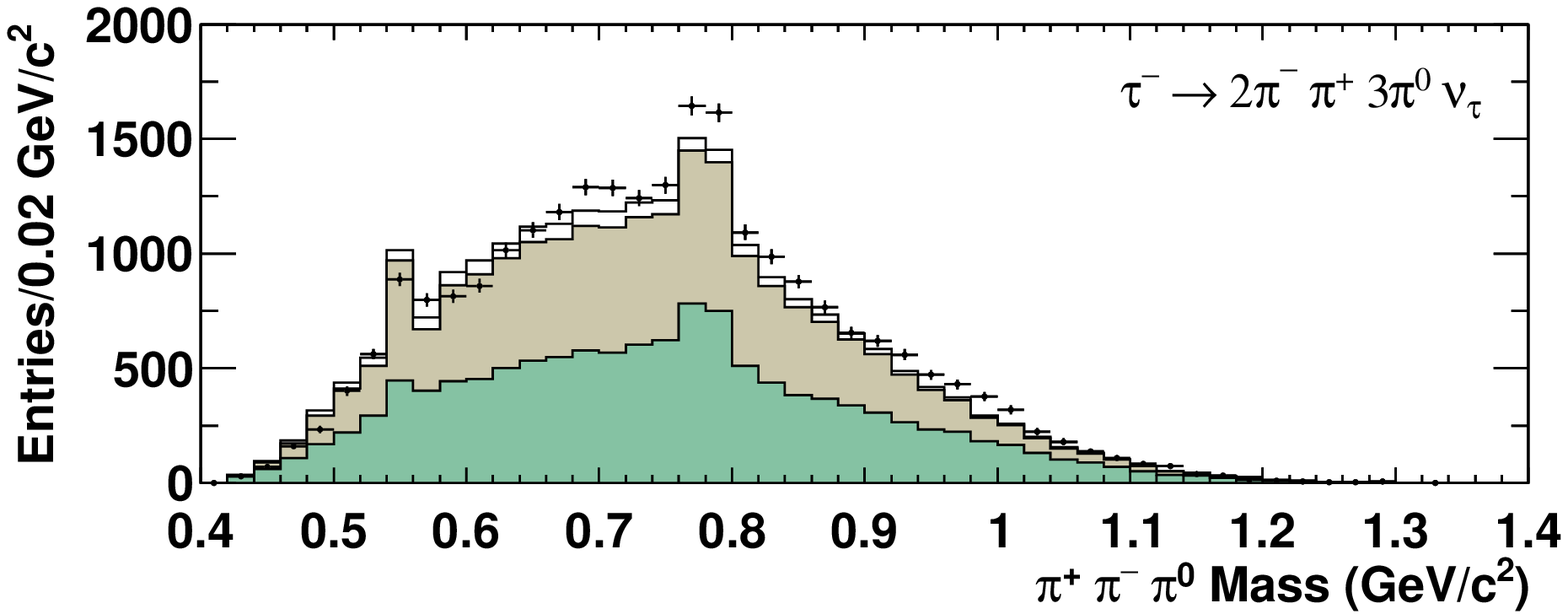, height=3.4cm}}
\mbox{\epsfig{file=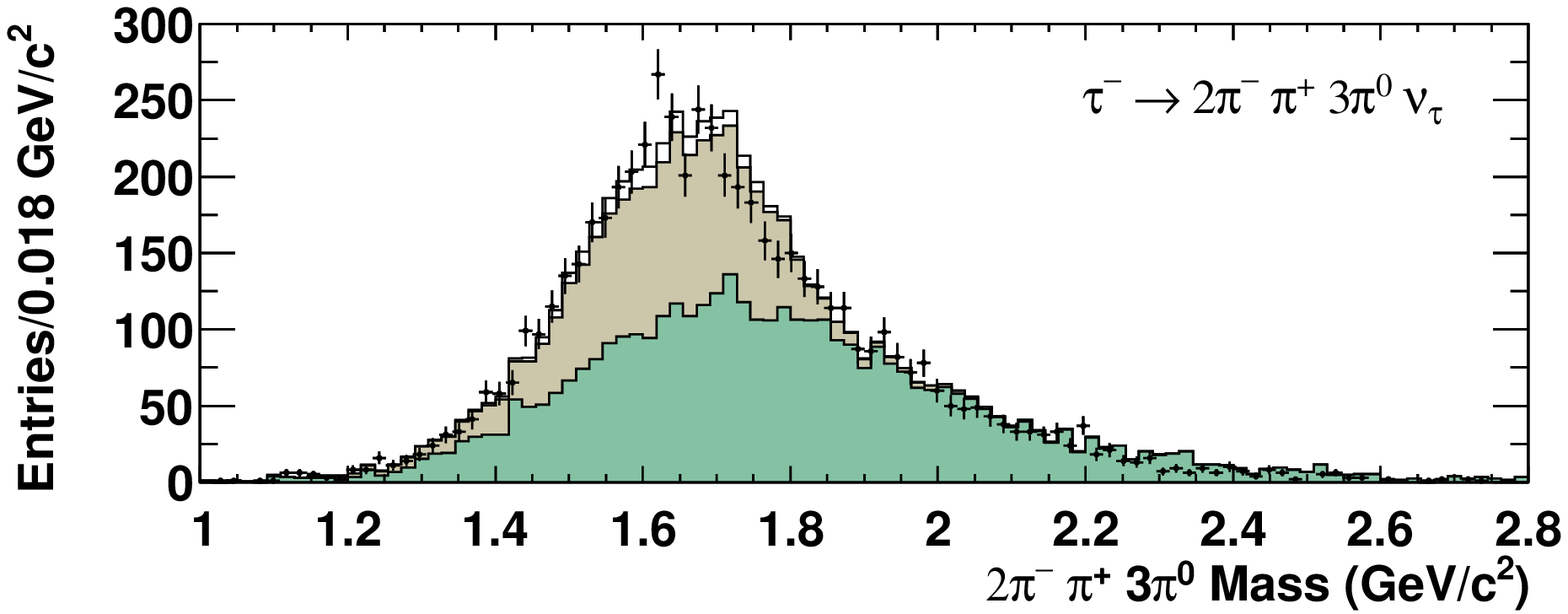,height=3.4cm}}
\caption{The 3\piz, $\pip\pim\piz$, and $2\pim\pip3\piz$
invariant mass distributions in \taupppzzz decay candidates.
The predictions of the Monte Carlo simulation are shown for the 
resonant and  non-resonant \mtau decays, and the background from other
\mtau decays and \qqbar events. 
The resonant decays include decays with the correct topology and a 
resonance ($\eta$, $f_1$ or $\omega$) in the final state.
}
\label{fig:3prong3pi0:plot}
\end{figure}

\begin{table*}[tbph]
\renewcommand\arraystretch{1.25}
\begin{center}
\caption{\label{table:omega} 
Results and branching fractions for \tauthreepiomega decays.}
\vspace{0.25cm}
\begin{tabular}{lcc} \hline \hline
Decay Mode   & \taupppo   & \taupzzomega \\
             &  \omegappp & \omegappp    \\
\hline 
Branching fractions ($10^{-5}$)   
                & \hspace{0.25cm} $\ombrbx$  \hspace{0.25cm} 
                & \hspace{0.25cm} $\BROmegaTwoPiPizCorrNoBra$ \hspace{0.25cm} \\
Data events          & $\omeganevt$
                     & $\NsigOmegaPiTwoPiz$ \\
$\chi^2/NDF$         & $\omegachisq/\omegandf$
                     & $\OmegaDataChiPiTwoPiz$ \\
Selection efficiency & $\omeffp$
                     & $\OmegaEffPiTwoPiz$ \\ 
Background           & $\omnqq$
                     & $\OmegaNbkgTotal$  \\
\hline 
Systematic uncertainties (\%)    & &   \\
Tracking efficiency  & \omsd 
                     & \OmegaPiTwoPiZeroTrackingSyst \\
\piz and $\eta$ PID  & \omsa
                     & \OmegaPiTwoPiZeroEtaEffSyst \\
Pion PID             & \omsb
                     & \OmegaPiTwoPiZeroPiPIDSyst \\
Lepton-tag PID       & \omsc 
                     & \OmegaPiTwoPiZeroPIDSyst \\ 
$N_{\tt}$            & \omse 
                     & \OmegaPiTwoPiZeroLumiSyst \\ 
Selection efficiency & \omsh
                     & \OmegaPiTwoPiZeroMCEffSyst \\
Background           & \omsg 
                     & \OmegaPiTwoPiZeroBkgSyst\\
$\mathcal{B}$(\omegappp)   & \omsf &  \OmegaPiTwoPiZeroBomegapipipizSyst \\
\hline 
Total (\%)           & \oms 
                     & \OmegaPiTwoPiZeroSystTotal \\ 
\hline \hline
\end{tabular}
\end{center}
\end{table*}

\begin{figure}[htpb]
\begin{center}
\mbox{\epsfig{file=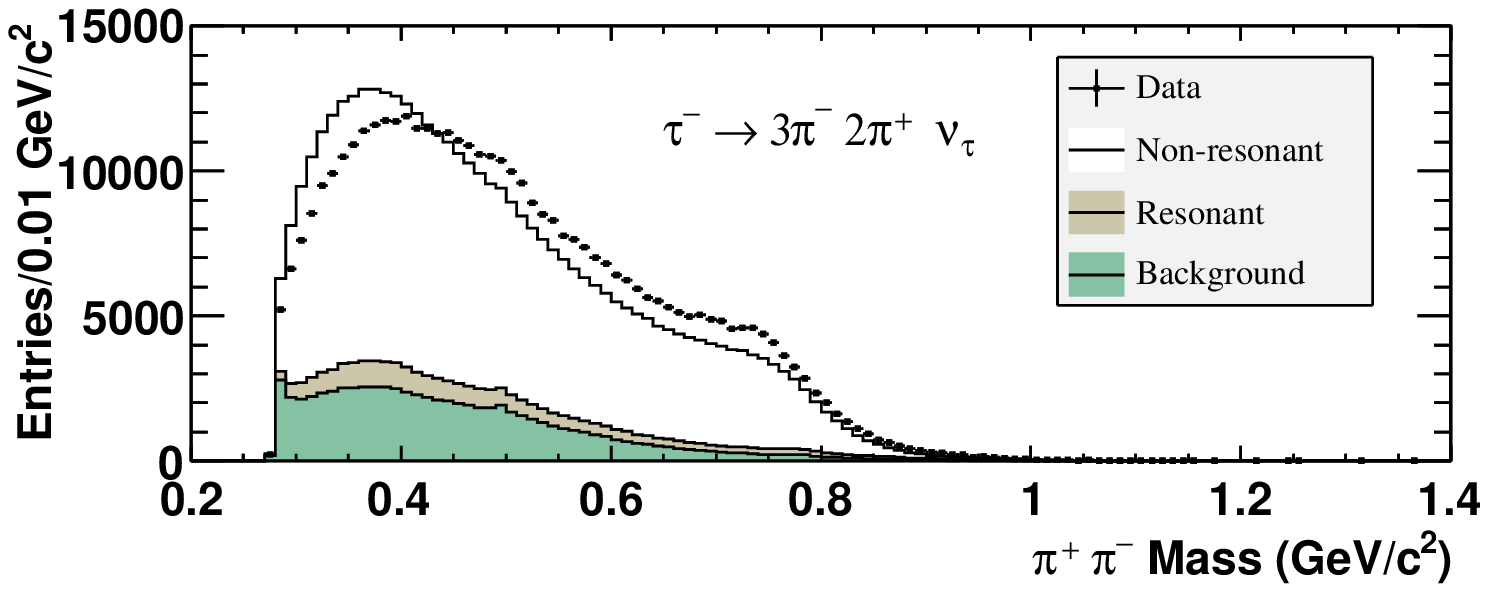,height=3.4cm}}
\mbox{\epsfig{file=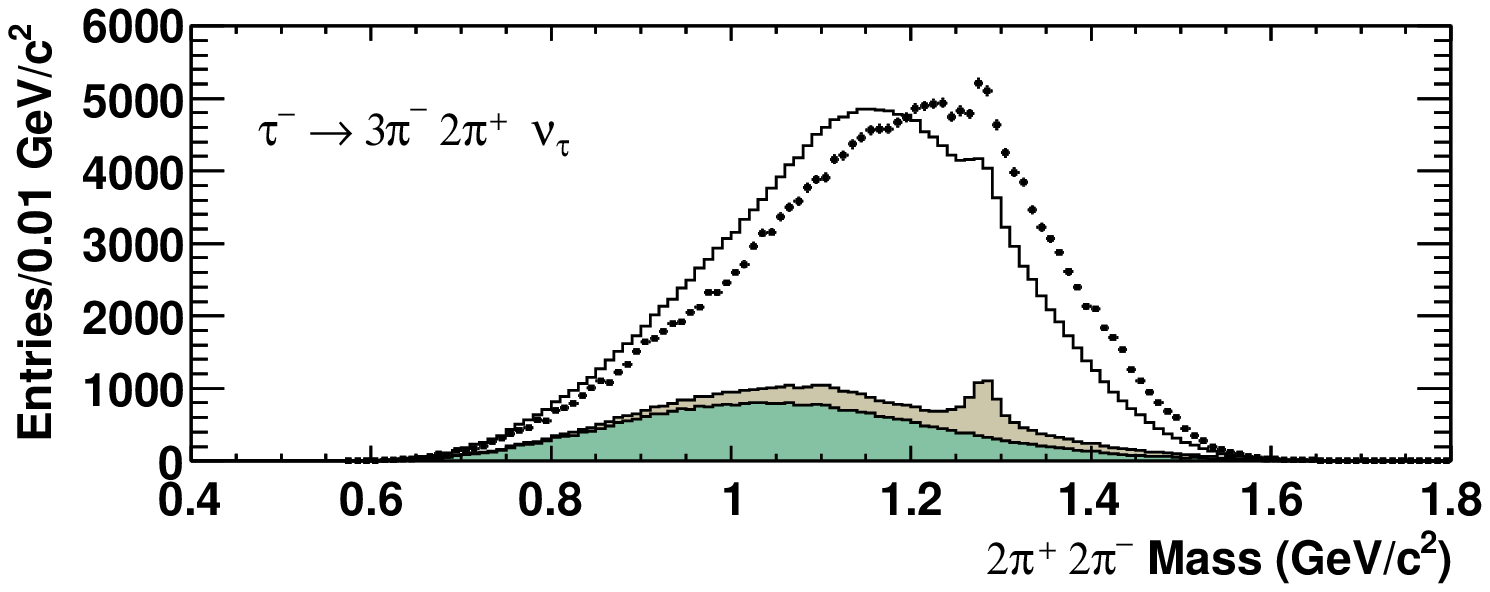,height=3.4cm}}
\mbox{\epsfig{file=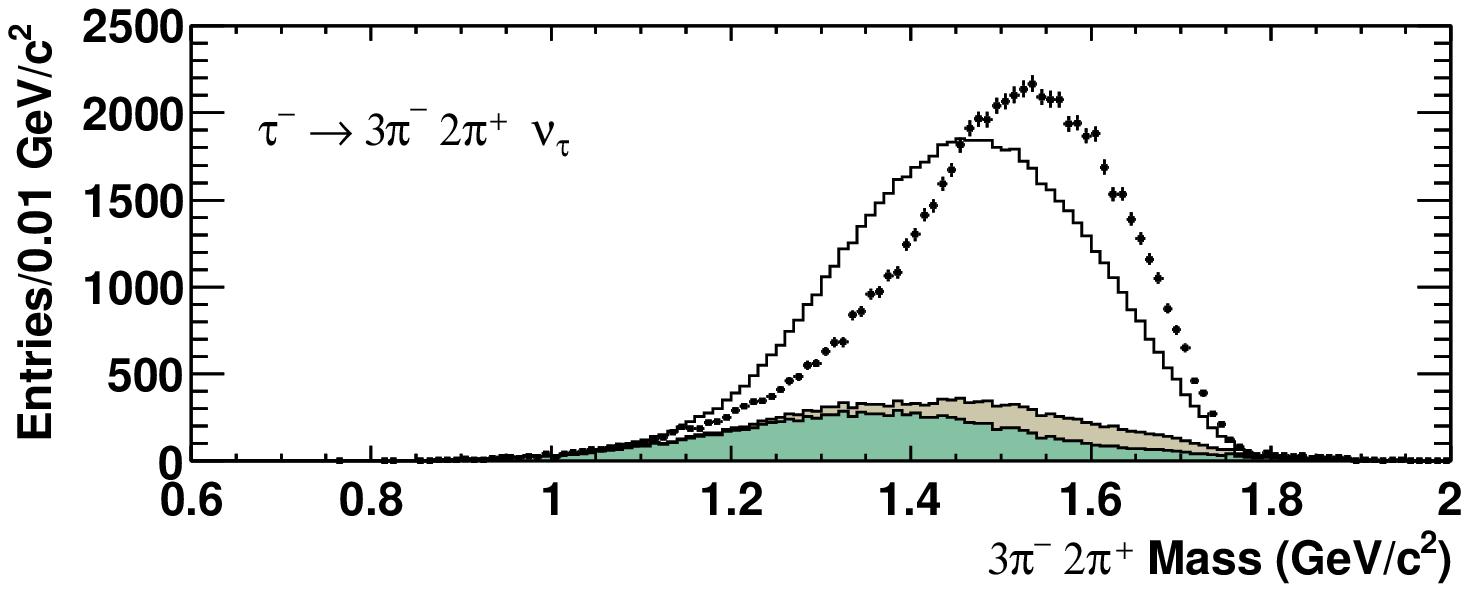,height=3.4cm}}
\end{center}
\caption{\label{fig:5pi}
The $\pip\pim$, $2\pip2\pim$, and $3\pim2\pip$ invariant mass distributions
in \taufivepi decays.  
The predictions of the Monte Carlo simulation are shown for the
resonant and  non-resonant \mtau decays, and the background from other
\mtau decays and \qqbar events.
The resonant decays include decays with the correct topology and a
resonance ($\eta$, $f_1$ or $\omega$) in the final state.
The non-resonant decays are generated using $\taum \rightarrow a_1^- \nut$ events.
The differences between the data and Monte Carlo predictions are
discussed in the text.
}
\end{figure}

\begin{figure*}[htpb]
\mbox{\epsfig{file=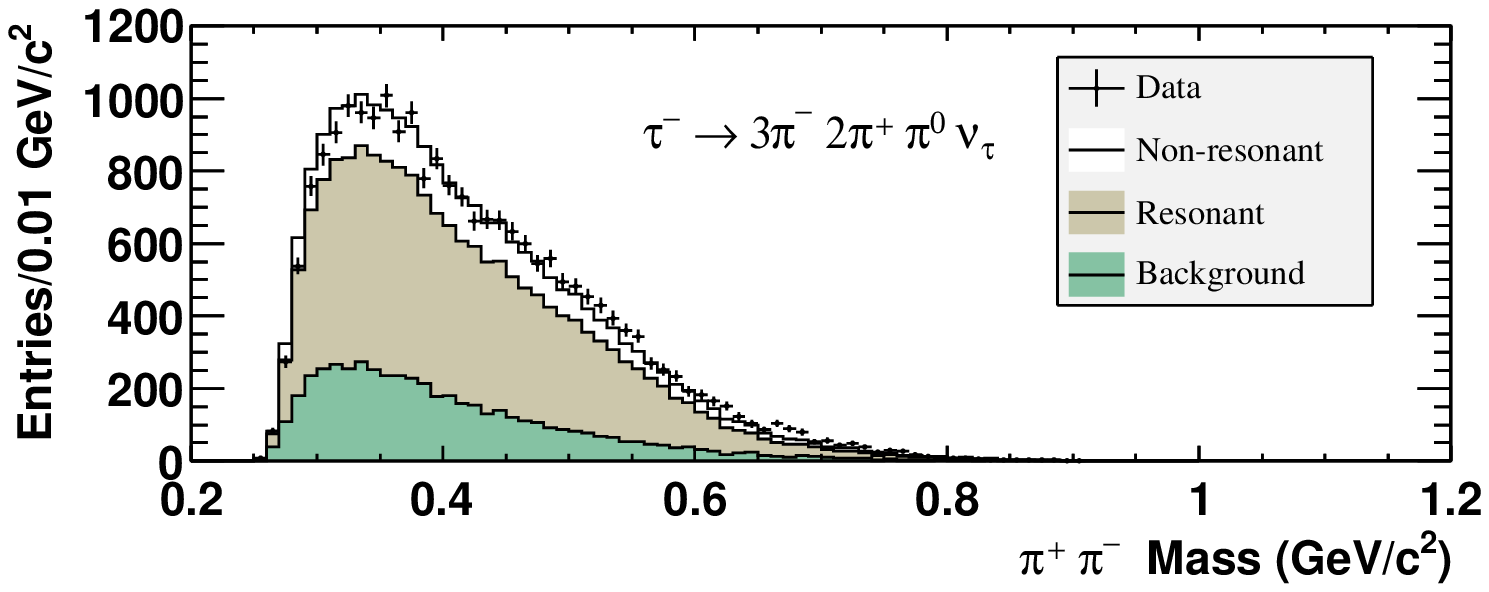,height=3.4cm}}
\mbox{\epsfig{file=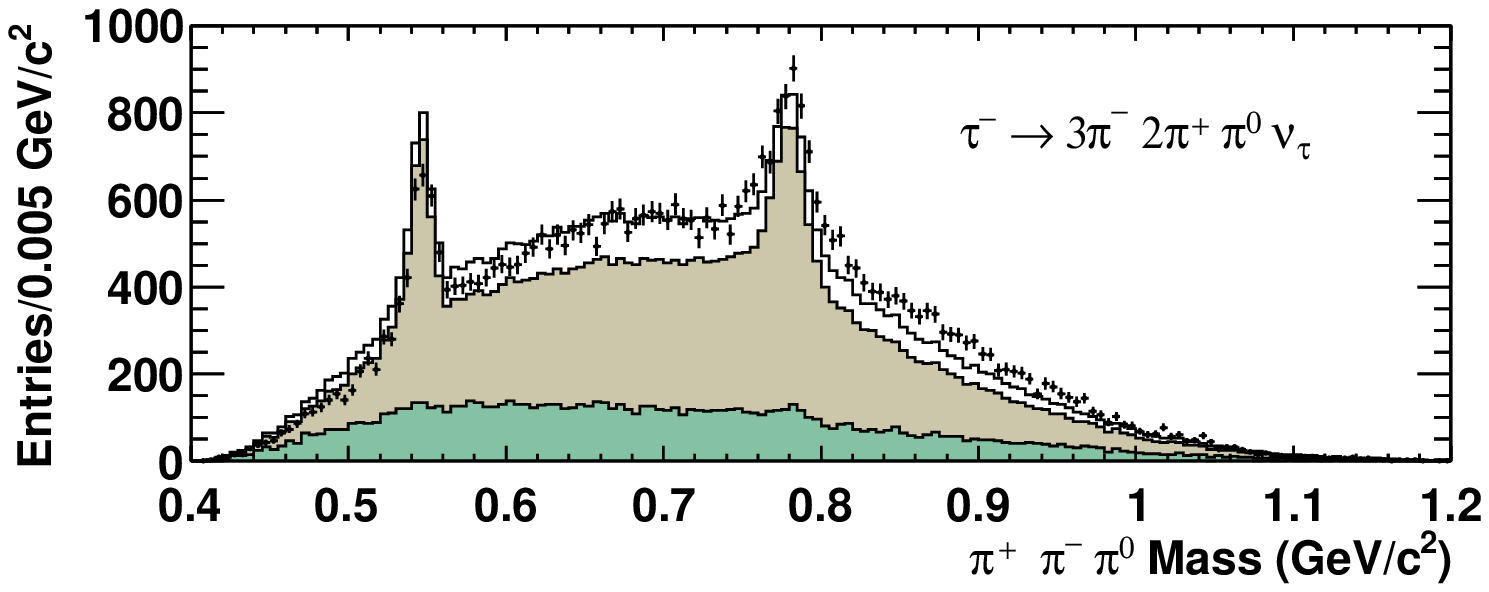,height=3.4cm}}
\mbox{\epsfig{file=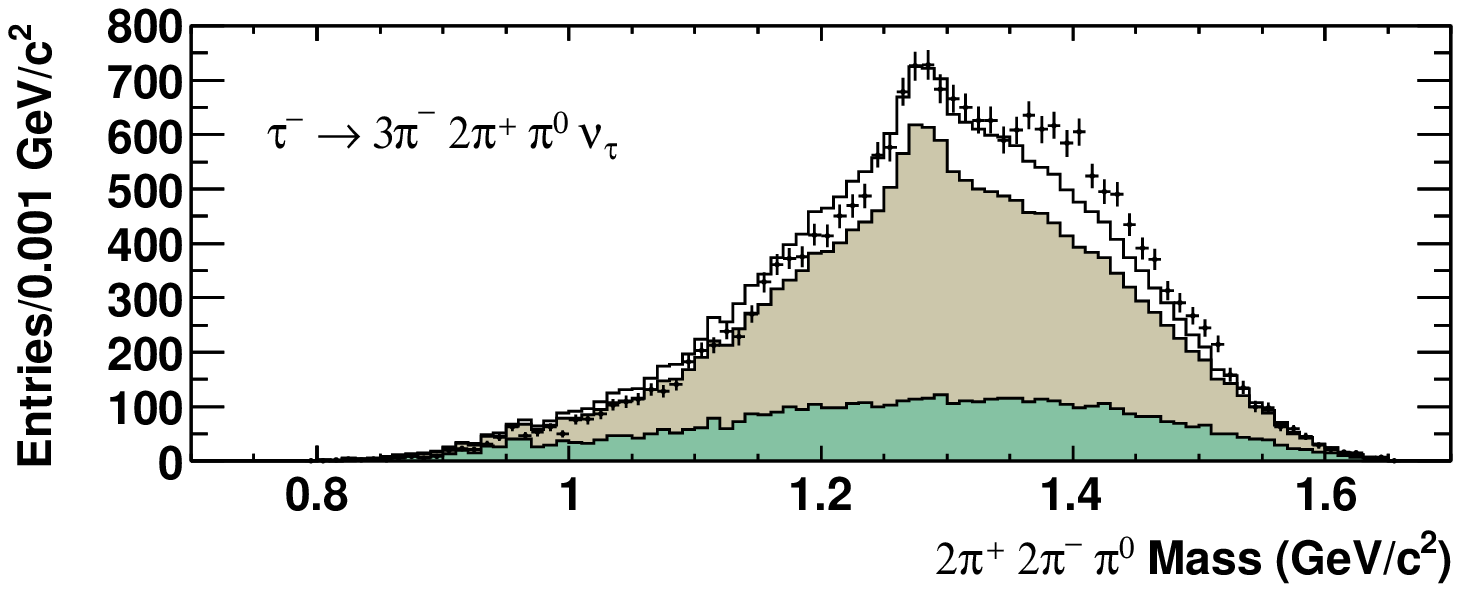,height=3.4cm}}
\mbox{\epsfig{file=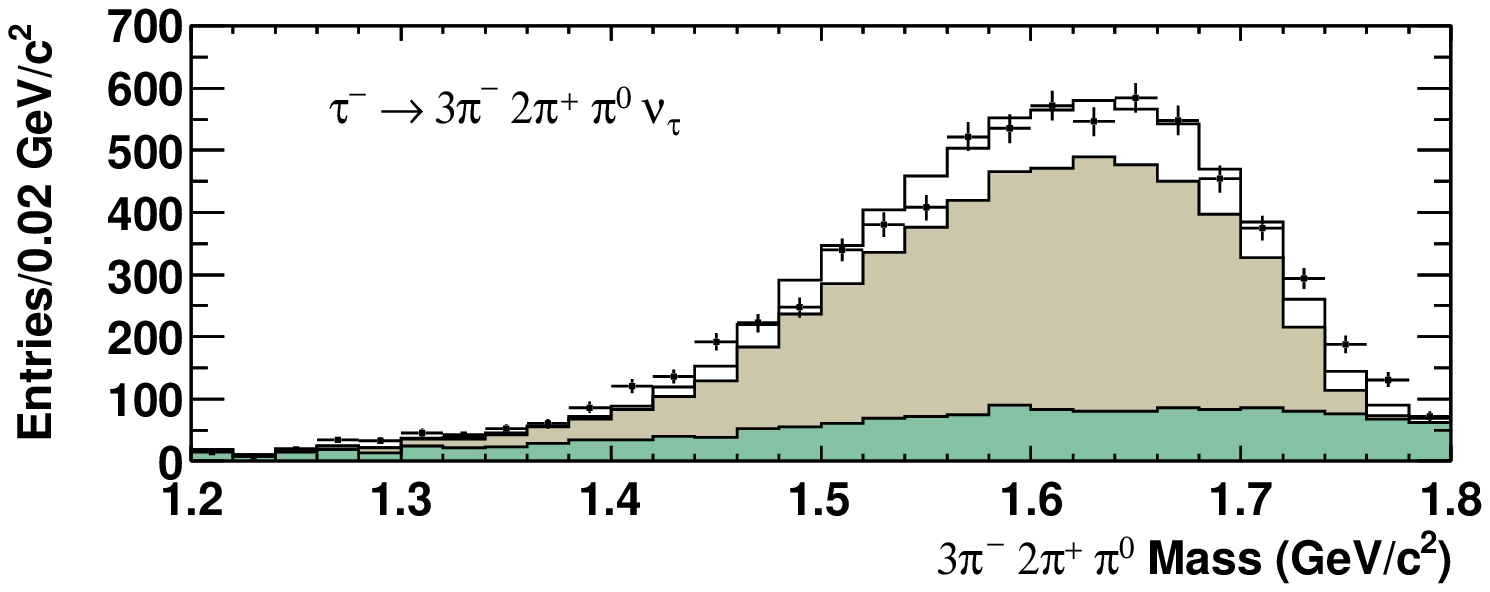,     height=3.4cm}}
\caption{\label{fig:6pi}
The $\pip\pim$, $\pip\pim\piz$, $2\pim2\pip\piz$,
and $3\pim2\pip\piz$ invariant mass distributions in
\taufivepipiz decays.
The predictions of the Monte Carlo simulation are shown for the
resonant and  non-resonant \mtau decays, and the background from other
\mtau decays and \qqbar events.
The resonant decays include decays with correct topology and a
resonance ($\eta$, $f_1$ or $\omega$) in the final state.
}
\end{figure*}

\subsection{\tauthreepiomega}

We measure the \taupppo and \taupzzomega branching fractions.
The number of events is determined by fitting the $\omega$ peak in the
$\pip\pim\piz$ invariant mass distributions (see Fig.~\ref{fig:omega})
with a Breit-Wigner distribution, which is convolved 
with a Gaussian distribution to take into account the detector resolution.
The resolution parameter of the Gaussian distribution is determined using 
a data control sample consisting of \qqbar events, and is fixed in the fit.
A polynomial function is used to fit the background.
The results are presented in Table~\ref{table:omega}.

Approximately 10\% of the events in the \taupppo channel are 
backgrounds from other \mtau decays (primarily 
$\taum \! \rightarrow  \pim\piz\omega \nut$ decays)
and $\epem \rightarrow \qqbar$ events. 
The backgrounds are subtracted before calculating the branching fraction. 

The \taupzzomega sample has substantial contributions from
$\taum \! \rightarrow  \pim\omega \nut$ and
$\taum \! \rightarrow  \pim\piz\omega \nut$ decays.
The background is estimated with the Monte Carlo simulation and verified 
using data and simulation control samples.
The control samples follow the nominal selection criteria but select
one or two \piz instead of three \piz mesons.

The branching fractions  are found to be  
\begin{eqnarray*}
\mathcal{B}(\taupppo)     & = \ombrb, \\
\mathcal{B}(\taupzzomega) & = \BROmegaTwoPiPizCorr .
\end{eqnarray*}
The systematic uncertainties are listed in Table~\ref{table:omega}.

The \taupppo and \taupzzomega branching fractions are consistent with the 
results from CLEO,  $(1.2 \pm 0.2 \pm 0.1) \times 10^{-4}$  and
$(1.4 \pm 0.4 \pm 0.3) \times 10^{-4}$, respectively \cite{cleo:sixpi}.
Gao and Li suggest that this mode is dominated by the ($\pi \rho \, \omega$) state
and predict a branching fraction in the range of  $1.8-2.1 \times 10^{-4}$
with the two modes (\taupppo and \taupzzomega) having the same value
\cite{theory:gao}.
The result measured in this work is approximately 50\% of the predicted rate 
but the ratio of the two branching fractions is consistent with unity.

\begin{figure}[tbph]
\begin{center}
\mbox{\epsfig{file=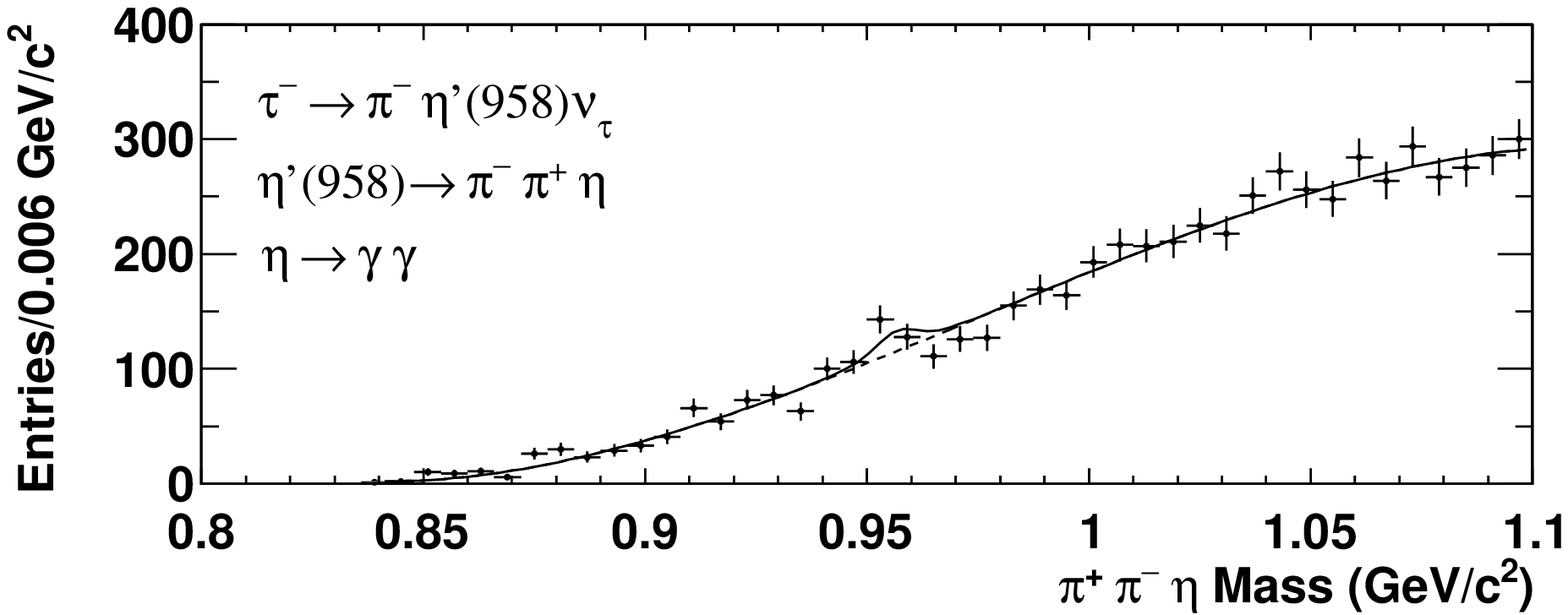,       height=3.4cm}}
\mbox{\epsfig{file=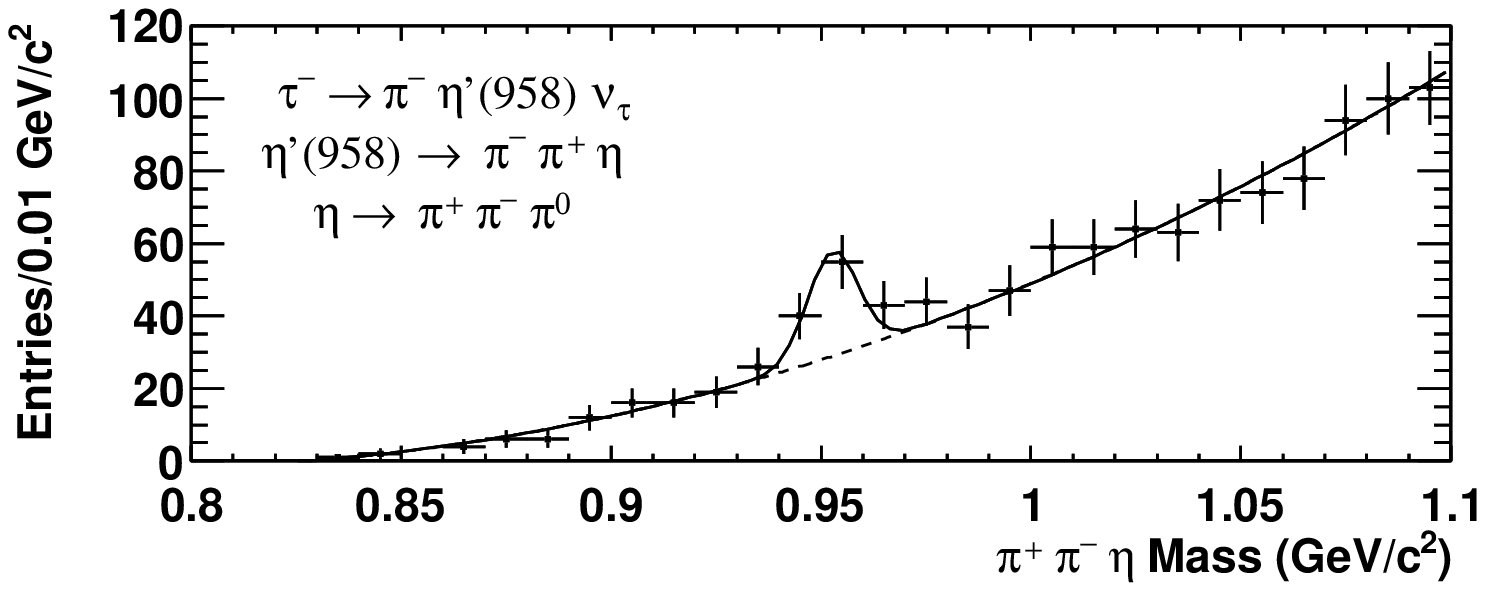, height=3.4cm}}
\mbox{\epsfig{file=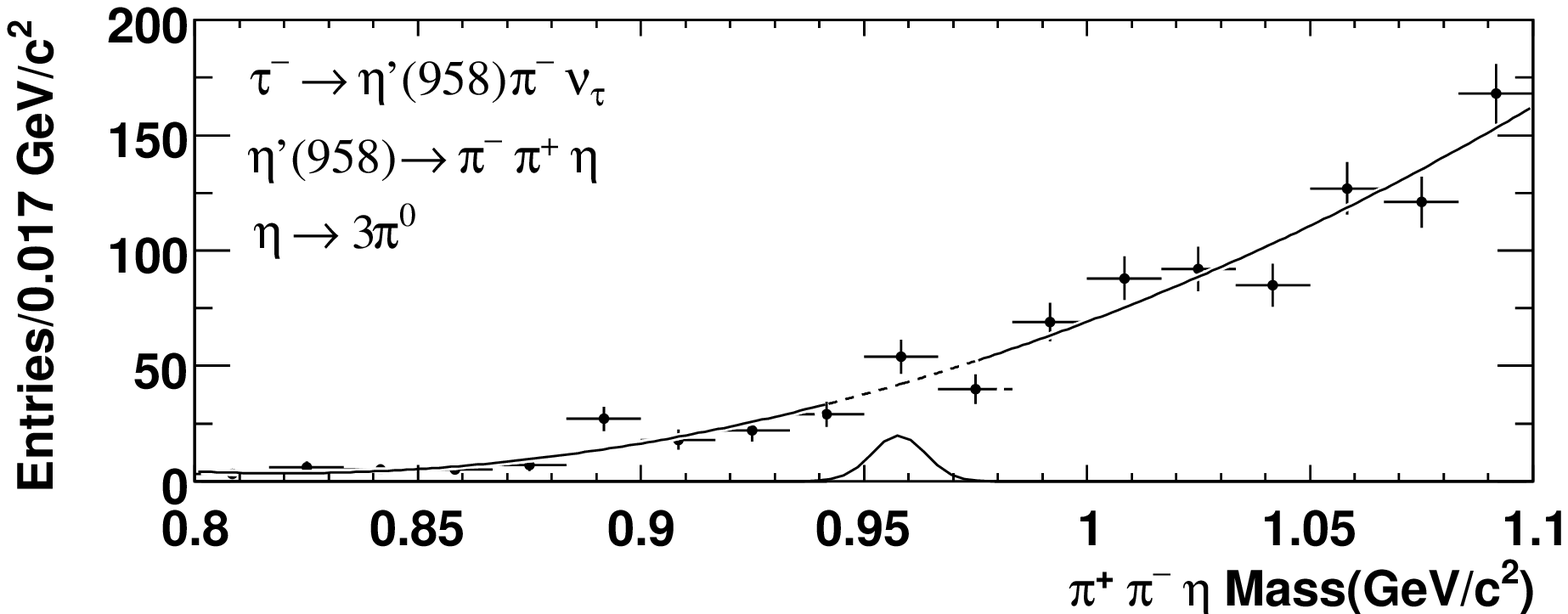,   height=3.4cm}}
\end{center}
\caption{\label{fig:scc}
The $\pip\pim\eta$ invariant mass in \taupppe decay candidates for the \etagg, 
\etappp,  and \etapiz decay modes after all selection
criteria are applied.
The fit to the \etapr peak (in the top two plots) is represented 
by the solid line.
The number of \etapr candidates in the \etapiz channel is determined
by counting the number of events in a single bin centered on the \etapr mass
and subtracting the combinatorial events.
The number of combinatorial events is determined by fitting the data
(the solid line in the bottom plot), excluding the data point near the \etapr peak.
The peak in this plot indicates the expected location and width of an \etapr signal.  
}
\end{figure}

\begin{table*}[tbph]
\renewcommand\arraystretch{1.25}
\begin{center} 
\caption{\label{table:5pi:nonres} 
Results and branching fractions for 
\taupppzzz, \taufivepi, and \taufivepipiz non-resonant  decays.}
\vspace{0.25cm} 
\begin{tabular}{lccc} \hline \hline
Decay Mode           & \taupppzzz & \taufivepi & \taufivepipiz \\
\hline 
Branching fractions  ($10^{-4}$)
                     & \hspace{0.25cm} $\BRPSIncThreePizNoBra$  \hspace{0.25cm}
                     & \hspace{0.25cm} $\fpbfx$  \hspace{0.25cm}
                     & \hspace{0.25cm} $\sixbrx$ \hspace{0.25cm} \\
Data events          & $\NsigIncThreePiz \;\;$
                     & $\fpdatax \;\;$
                     & $\sixnevts \;\;$    \\
Efficiency           & $\EffPSIncThreePiz$
                     & $\fpseleffp$ 
                     & $\sixEffp$ \\
\hline 
Background & & & \\
Resonant             & $\NResMCIncThreePiz$
                     & $\fpbfpi$
                     & $\sixbkgds$ \\
Other $\tau$ decays  & $\NOthertSum \;\;$
                     & $\fpbtau \;\;$
                     & $\sixbkgdt$ \\ 
\qqbar               & $\NQQIncThreePiz$
                     & $\fpbqq$
                     & $\sixbkgdq$ \\
\hline
Total Background     & $\NBkgPSIncThreePiz$
                     & $\fpbtot\;\;$
                     & $\sixbkgd$ \\
\hline
Systematic uncertainties (\%)    & & &  \\
Tracking efficiency  & 
                     & \fptk
                     & \sixsb \\
\piz and $\eta$ PID  & 
                     & -
                     & \sixse \\
Pion PID             & 
                     & \fppi
                     & \sixsd \\
Lepton-tag PID       & 
                     & \fplep
                     & \sixsc \\
\lum \sigtt          & 
                     & \fplumi
                     & \sixsa \\
Selection efficiency & 
                     & \fpeff
                     & \sixsh \\
Background           & 
                     & \fpbk
                     & \sixsi \\
\hline
Total (\%)           & 
                     & \fptot
                     & \sixsyst\\
\hline \hline
\end{tabular}
\end{center}
\end{table*}

\subsection{Non-resonant decay modes}

The resonant modes, involving $\eta$, $\omega$ and $f_1$ mesons, do not 
account for all of the observed decays, as discussed below. 
We consider the excess in the observed
decays to be from ``non-resonant'' modes.
We make no attempt to identify the contribution of resonances
with larger widths ($ > 100 \mevcc$) as the nature of these
resonances is complex and their lineshapes will be modified by the 
limited phase space in the \mtau decay.
The Monte Carlo simulation describes the final states using a phase-space 
model for the final-state particles.
The only exception is the \taufivepi mode, which Tauola models 
using $\taum \rightarrow a_1^- \nut$  decays \cite{kuhnwas}.

We measure the branching fractions of the non-resonant \taupppzzz, \taufivepi,
and \taufivepipiz decays. 
The numbers of candidates are given by the numbers of events found in the data 
after subtracting the resonant contributions and the background from 
other \mtau decays and \qqbar events (see Table~\ref{table:5pi:nonres}).

The invariant mass plots in Fig.~\ref{fig:3prong3pi0:plot} show that the 
resonant decays dominate the \taupppzzz mode.
The background is primarily from 
$\taum \rightarrow  \pim\piz\omega \nut$ and \qqbar events.
The branching fraction of the non-resonant \taupppzzz channel is determined to be 
$\BRPSIncThreePiz$.
The systematic uncertainty on the branching fraction is dominated by the 
uncertainty in the background, which includes the Monte Carlo statistical 
uncertainty and the \mtau branching fraction uncertainties.
We do not list the fractional systematic uncertainties for this mode 
in Table~\ref{table:5pi:nonres} due to the smallness of the branching fraction.
The branching fraction is consistent with zero and we set a limit of 
\begin{eqnarray*}
\mathcal{B}(\taupppzzz)  & < \CLIncThreePiThreePiz 
\end{eqnarray*}
at the 90\% confidence level.

We also determine the inclusive \taupppzzz branching fraction, 
given by the sum of the resonant and non-resonant terms.  We find the result 
$\BRTotalIncThreePiz$,
where the systematic uncertainty accounts for correlations 
between the systematic uncertainties of the individual modes.

The \taufivepi decay has only a small
contribution from resonant decays (see Fig.~\ref{fig:5pi}).
The branching fraction of the non-resonant \taufivepi decay is determined to be 
\begin{eqnarray*} 
\mathcal{B}(\taufivepi) & =  \fpbf \times 10^{-4} .
\end{eqnarray*}
The \taupppo (\omegappg) decay is considered as a resonant background and is not
included in the non-resonant branching fraction.
Although the modeling of the $3\pim 2\pip$ invariant mass distribution
is deficient, the selection efficiency remains the same if the 
Monte Carlo is re-weighted to resemble the data distribution.
The decay model represents a significant improvement compared to a 
phase-space model, in which the $\rho$ meson, observed in the 
$\pip\pim$ invariant mass spectrum, is not included.
Further tuning of the model is required, which is outside the scope
of the present study.
The background from the \qqbar events is validated by comparing the numbers
of data and Monte Carlo events in the region above the $\tau$ lepton mass.
The inclusive \taufivepi branching fraction is
$\fpbfinc$$\times 10^{-4}$ and is obtained by adding the non-resonant branching fraction
with the resonant branching fraction for the \taufpi via \fpppp decay. 
The branching fraction of the \taufiveh decay (where $h^-$ is either a \pim or \Km) 
was measured to be $(8.56 \pm 0.05 \pm 0.42) \times 10^{-4}$ 
in a previous \babar\ analysis \cite{babar:5prong} 
using a smaller data sample, which did not use charged particle identification
and in which the branching fraction included the contribution of the \taupppo decay.

\taufivepipiz decays are dominated by the resonant modes (see Fig.~\ref{fig:6pi}). 
We determine the
branching fraction of the non-resonant \taufivepipiz decay mode to be 
\begin{eqnarray*}
\mathcal{B}(\taufivepipiz) & = \sixbr .
\end{eqnarray*}
The systematic uncertainty on the non-resonant \taufivepipiz branching fraction 
is dominated by the large uncertainty in the background 
(see Table~\ref{table:5pi:nonres}).
Although the invariant mass distributions of the resonant modes in the Monte Carlo
simulation are corrected to provide better agreement with the data, 
the corrections make little difference to the final branching fraction result.
The other $\tau$ decays and the \qqbar events contribute to a lesser extent;
their contribution to the uncertainty of the background is very small.

The \taufivepipiz (including $\omega$ and excluding $\eta$)
branching fraction is  $\sixbriso$ and is obtained by adding
the non-resonant branching fraction and the resonant branching fraction
attributed to the \taupppo via \omegappp decay.
In addition, the inclusive \taufivepipiz branching fraction is
$\sixbrsum$ and is similarly obtained by adding 
the non-resonant branching fraction and the resonant branching fractions
attributed to the \taupppe via \etappp and \taupppo via \omegappp decays.

The \taufivepipiz (including $\omega$ and excluding $\eta$)
branching fraction can be compared with isospin model predictions
\cite{theory:pais, theory:sobie}.
There are three $\tau$ decay modes with six pions in the final state:
\taupppzzz, \taufivepipiz, and \taupzzzzz (there are no measurements of 
the \taupzzzzz decay mode).
There are four possible isospin states for six pion decays: ($4\pi \rho$), 
($3\rho$), ($3\pi \omega$), and ($\pi \rho \omega$).
The relative rates of the decays can be used to identify the dominant
isospin states.
The approximate equality of the \taupppzzz and \taufivepipiz branching fractions 
suggest that the ($4\pi \rho$) and ($\pi \rho \omega$) modes should dominate.
The limited phase space imposed by the \mtau mass suppresses the higher mass states
and as a result we do not observe evidence of the $\rho$ meson in these decays.

\begin{figure*}
\begin{center}
\mbox{\epsfig{file=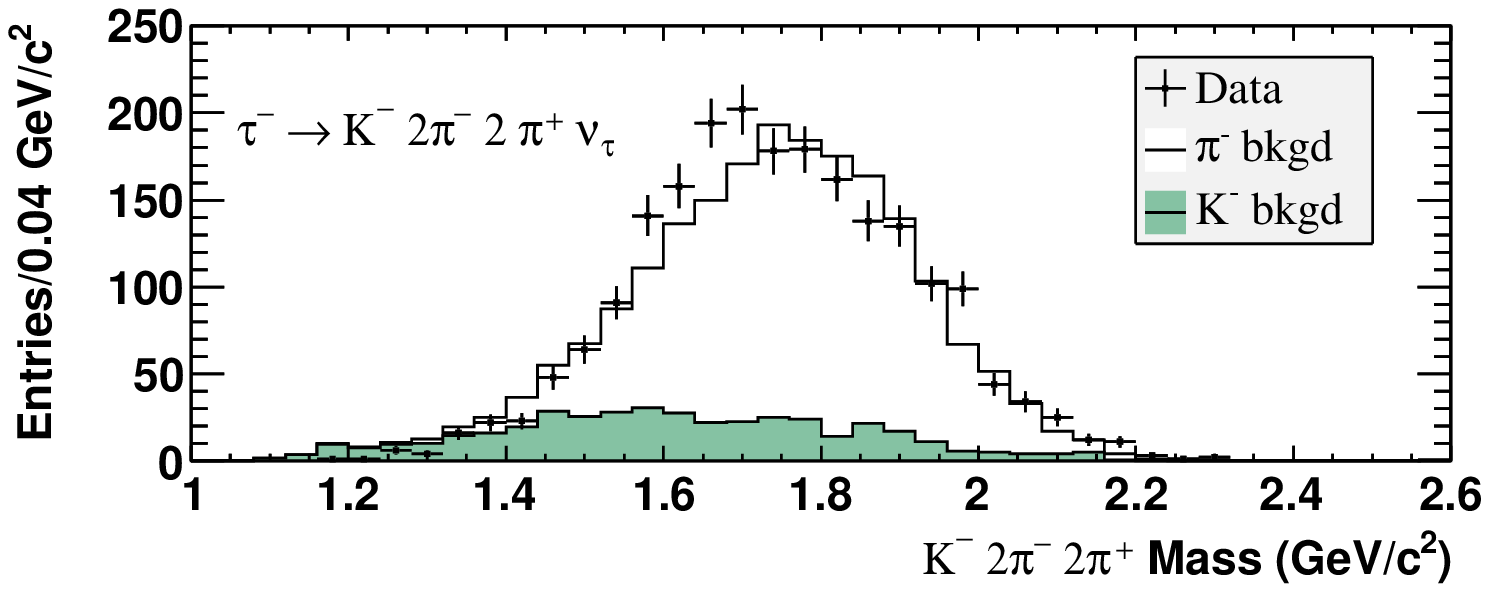,    height=3.2cm}}
\mbox{\epsfig{file=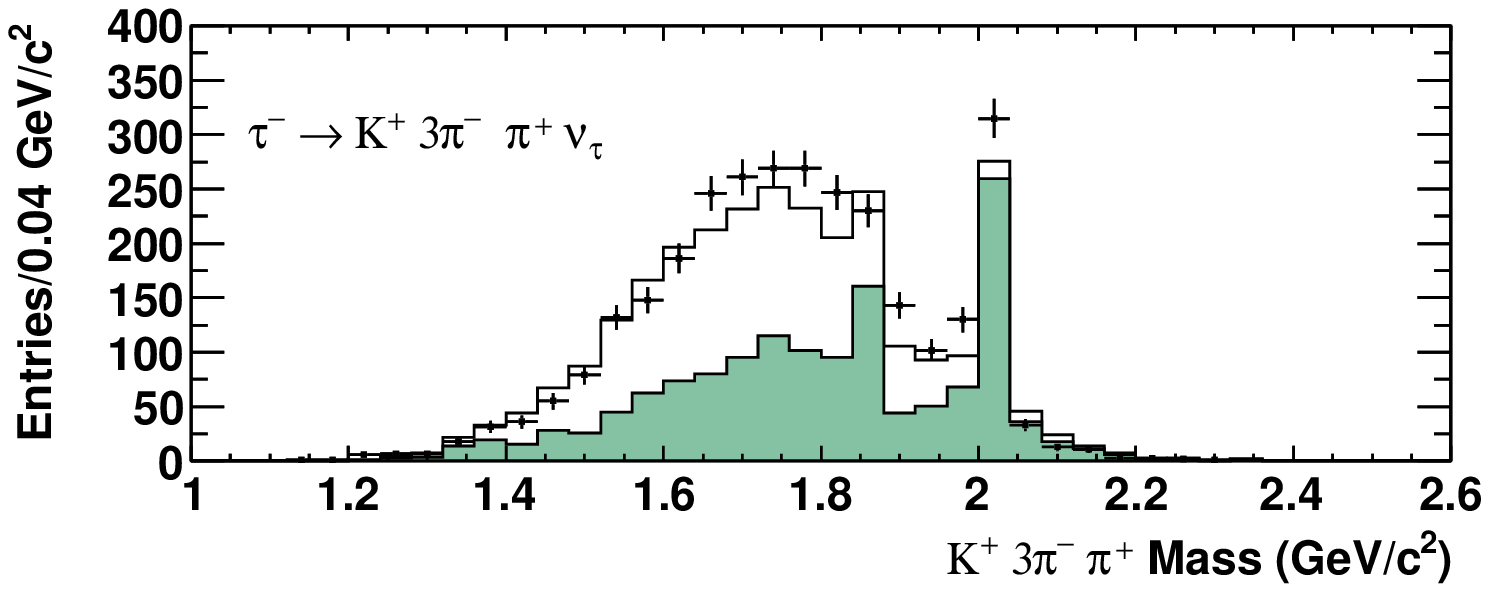,height=3.2cm}}
\mbox{\epsfig{file=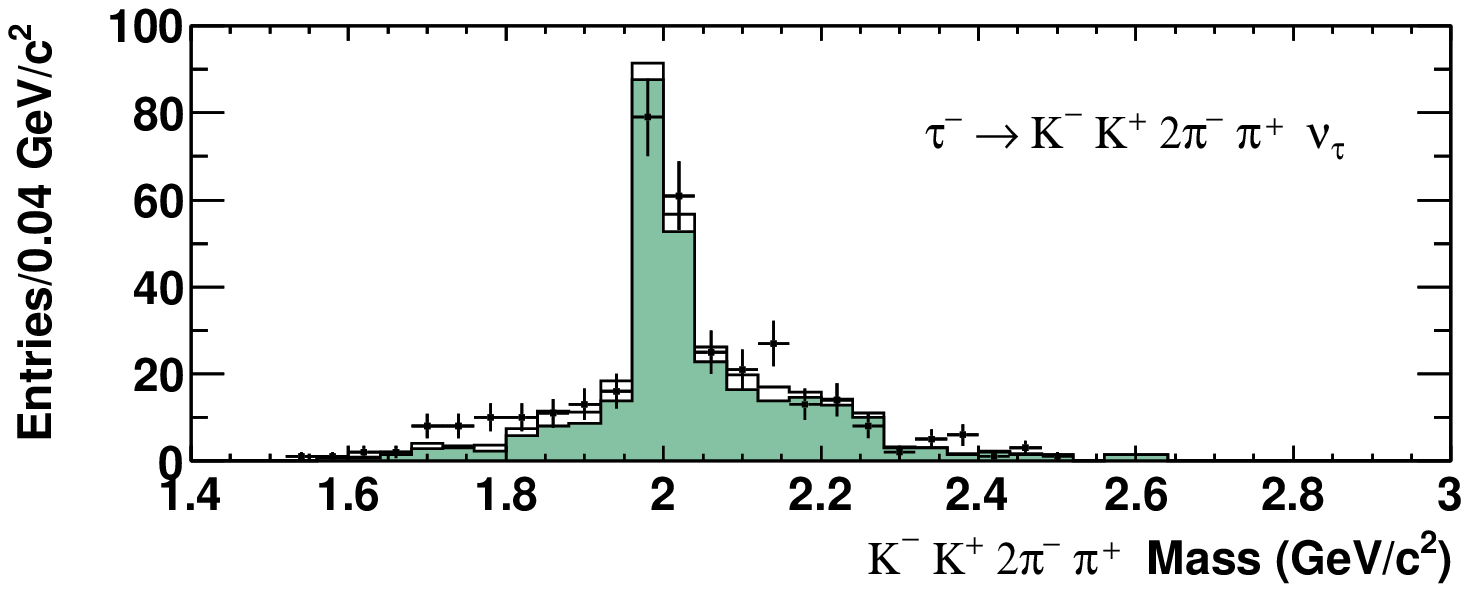,   height=3.2cm}}
\mbox{\epsfig{file=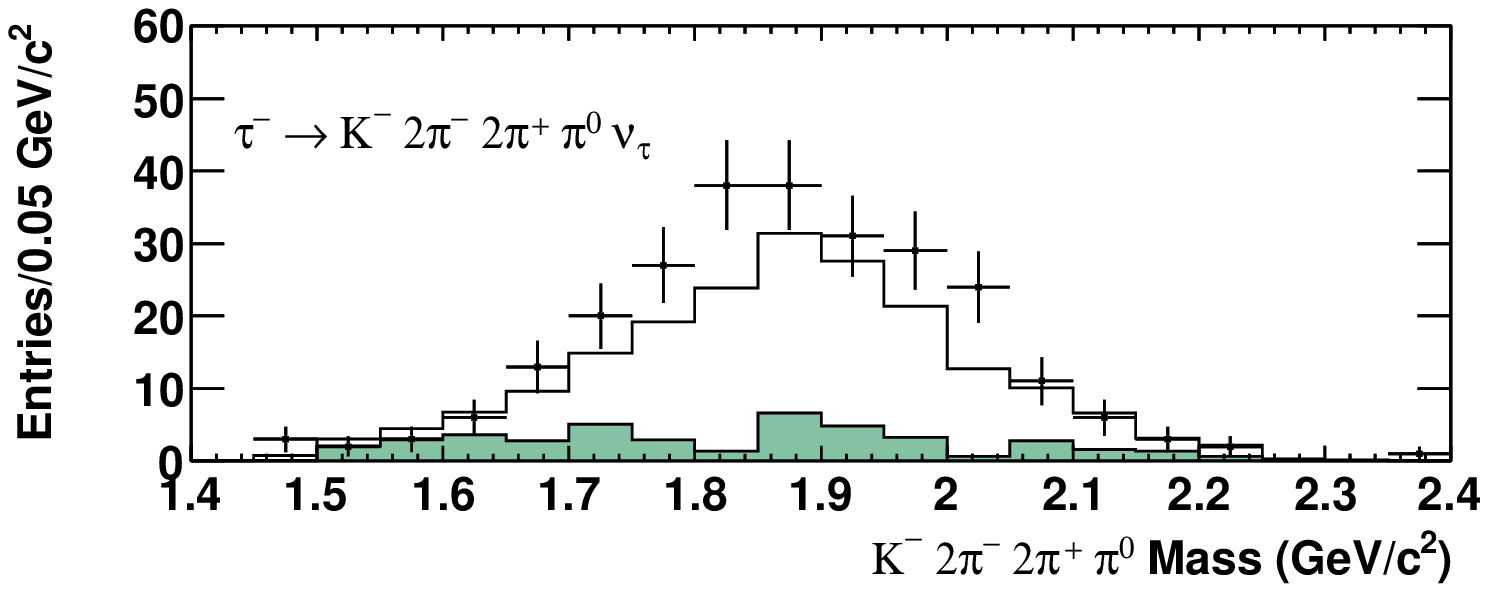,    height=3.2cm}}
\mbox{\epsfig{file=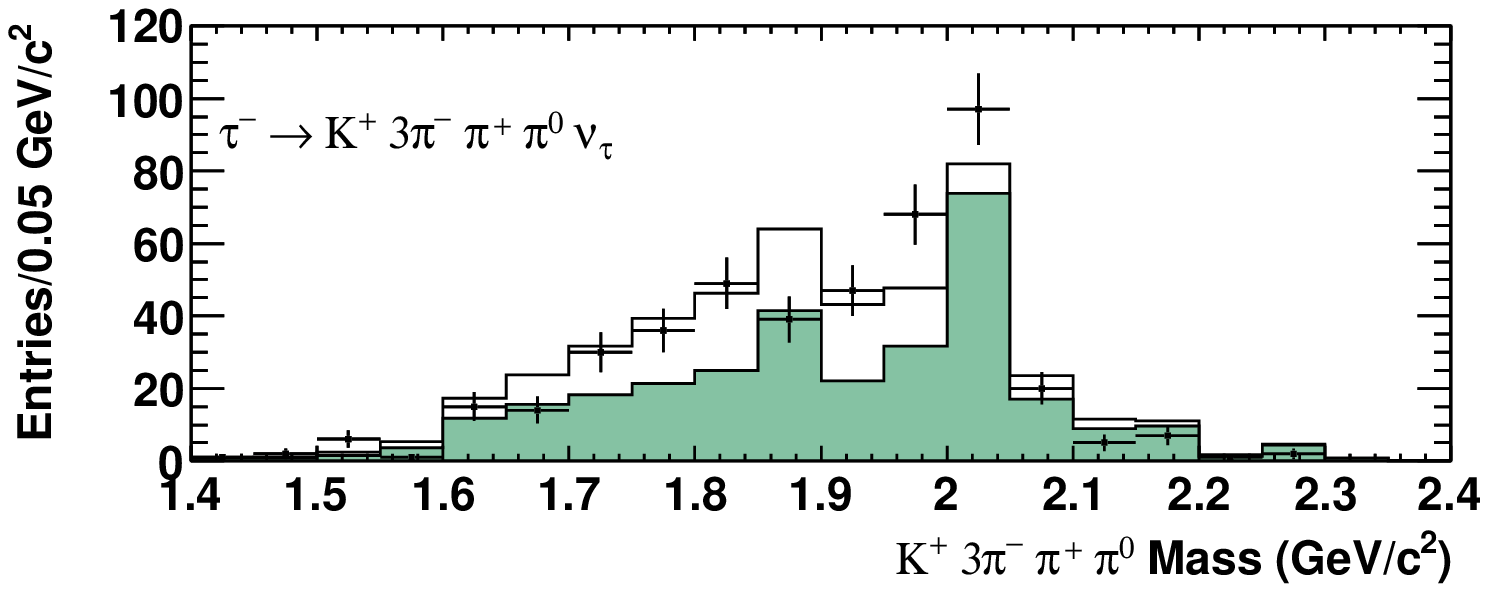,height=3.2cm}}
\end{center}
\caption{\label{fig:kaon}
The $\Km2\pim2\pip$, $\Kp3\pim\pip$, $\Km\Kp2\pim\pip$, 
$\Km2\pim2\pip\piz$, and $\Kp3\pim\pip\piz$ invariant mass distributions
in the data sample after all selection criteria are applied.
The unshaded histogram represents \mtau decays in which a charged pion 
is misidentified as a charged kaon, and the shaded histograms are primarily
from \qqbar events in which there is a charged kaon in the final state. 
The Monte Carlo simulation does not include signal decays.
}
\end{figure*}

\begin{table*}[tbph]
\renewcommand\arraystretch{1.25}
\begin{center}
\caption{\label{table:etap}
Results and branching fractions for 
\taupietaprimepiz, \tauketaprime, and \tauetaprime  decays.
}
\vspace{0.25cm}
\begin{tabular}{lccc} \hline \hline
& & & \\
\taupietaprimepiz  \hspace{1.5cm}
   & \hspace{0.5cm} \etagg  \hspace{0.5cm}
   & \hspace{0.5cm} \etappp \hspace{0.5cm}  & \\
\hline
Limit (90\% C.L.)                &  $\CLCorrEtaPrimePiz$       & $\pietappizxlimit$ &  \\
Branching fraction ($10^{-6}$)   &  $\BRCorrEtaPrimePizNoBra$  & $\pietappizxbra$   &  \\
Data events                      &  $\NdatEtaPrimePiz$         & $\pietappizxdata$  &  \\
Background events                &  $\NqqEtaPrimePiz$          & $\pietappizxbkgd$  &  \\
Selection efficiency             &  $\EffEtaPrimePiz$          & $\pietappizxeff$   &  \\
\hline
& & & \\
\tauketaprime

   & \hspace{0.25cm} \etagg  \hspace{0.25cm}
   & \hspace{0.25cm} \etappp \hspace{0.25cm}  & \\
\hline
Limit (90\% C.L.)                &  $\CLCorrEtaPrimeKaon$       & $\ketaplimit$ &  \\
Branching fraction ($10^{-6}$)   &  $\BRCorrEtaPrimeKaonNoBra$  & $\ketapbra$   &  \\
Data events                      &  $\NdatEtaPrimeKaon$         & $\ketapdata$  &  \\
Background events                &  $\NqqEtaPrimeKaon$          & $\ketapbkgd$  &  \\
Selection efficiency             &  $\EffEtaPrimeKaon$          & $\ketapeffx$  &  \\
\hline
& & & \\
\tauetaprime
   & \hspace{0.25cm} \etagg  \hspace{0.25cm}
   & \hspace{0.25cm} \etappp \hspace{0.25cm}
   & \hspace{0.35cm} \etapiz \hspace{0.25cm} \\
\hline
Limit (90\% C.L.)     & $\PrimeGGCL$
                      & $\sccxblimitb$
                      & $\PrimeThreePizCL$ \\
Branching fraction ($10^{-6}$)
                      & $\PrimeGGBRNoBra$
                      & $\sccxbrcx$
                      & $\PrimeThreePizBRNoBra$ \\
Data events           & $\PrimeGGNsig$  & $\sccxndata$ & $\PrimeThreePizNsig$ \\
Background events     & $\PrimeGGNqq$   & $\sccxnmc$   & $\PrimeThreePizNqq$  \\
Selection Efficiency              & $\PrimeGGEff$   & $\sccxeffp$  & $\PrimeThreePizEff$  \\
\hline \hline 
\end{tabular}
\end{center}
\end{table*}

\subsection{Search for decays involving \etapr(958) decays }

We also search for the \taupietaprimepiz, \tauketaprime, and \tauetaprime 
decays, where \etap.
The first two decays are allowed first-class-current decays whereas the 
last decay is a second-class-current decay, with a rate that
would be zero in the limit of perfect isospin symmetry.

The numbers of \etapr candidates in the data and background Monte Carlo 
samples are given in Table~\ref{table:etap}.
For the \taupietaprimepiz via \etagg and the
\tauetaprime via \etagg and \etappp modes, we measure the number of
\etapr candidates by fitting the peak with a Gaussian 
function and the combinatoric background with a polynomial function.
The number of \etapr candidates in the other channels is determined
by counting the number of events in a single bin centered on the \etapr mass
and subtracting the combinatorial events.
The level of the combinatorial background is estimated by fitting the 
mass spectrum or from the average level of the sideband region around
the \etapr peak.

The $\pip\pim\eta$ invariant mass distributions for the \tauetaprime
candidate decays are shown in Fig.~\ref{fig:scc}.
Although we see an \etapr peak in the \etappp channel, we find that it can
be fully accounted for by \qqbar events.
We do not show the invariant mass distributions for the 
\taupietaprimepiz and \tauketaprime decays.
The analysis of these decay modes uses only the \etagg and \etappp channels. 
The \etapiz channel was not considered due to the limited size of the samples.

The results for the three decay modes are given in Table~\ref{table:etap}.
The selection efficiencies are determined with the signal Monte Carlo samples.
The backgrounds from \etapr mesons are attributed to \epem $\rightarrow$ \qqbar 
events and estimated using the Monte Carlo samples.
The background estimations are validated by comparing the prediction of the
Monte Carlo simulation with data for events where the invariant mass of all
the observed final-state particles is greater than the \mtau mass.

We find no evidence for \taupietaprimepiz, \tauketaprime, or \tauetaprime decays
(see Table~\ref{table:etap}) and place the following upper limits 
on the branching fractions at the 90\% confidence level:
\begin{eqnarray*}
\mathcal{B}(\taupietaprimepiz) & < \pietaprimepizlimit, \\ 
\mathcal{B}(\tauketaprime)     & < \ketaprimelimit, \\
\mathcal{B}(\tauetaprime)      & < \sumpietaprimelimitb.  
\end{eqnarray*}
The limits are determined from the weighted average of the branching fractions
measured for each mode.
The \taupietaprimepiz and \tauketaprime channels are potential backgrounds to the
\tauetaprime decay. 
We find that background from these two decays is less 
than two events based on the upper limits on the branching fractions 
and we consider these backgrounds to be negligible.
The previous limits on the \tauetaprime decay were measured by 
\babar\ to be $7.2 \times 10^{-6}$ \cite{babar:3pieta} and by 
CLEO to be $8\times 10^{-5}$ \cite{cleo:3pieta}.
It is predicted that the branching fraction of
\tauetaprime should be less than $1.4 \times 10^{-6}$ \cite{nussinov}.


\subsection{Searches for decays involving charged kaons}
Finally we present the first search for high-multiplicity \mtau decays with one 
or two charged kaons.
We find no evidence for signal decays and place upper limits on the 
branching fractions of the 
\taukfourpi, \taukoppfourpi, \taukkppp, \taukfourpipiz, 
\taukoppfourpipiz, and \tauketaprime decay modes (the \tauketaprime decay
was presented in Section~E).

The events are divided into topologies in which the charged kaon has either
the same or opposite charge as the parent $\tau$ lepton.
If there are two kaon candidates, they must have opposite charge.
All other tracks are required to be identified as charged pions.
The selection criteria and systematic uncertainties are described earlier.
The requirement on the invariant mass ($M < 1.8 \gevcc$) of the final 
state uses the kaon mass for tracks identified as charged kaons.
Figure~\ref{fig:kaon} shows the mass spectra for the various channels.
The predictions of the Monte Carlo simulation are divided into decays
with or without a \Km (in this latter case, a \pim is misidentified 
as a \Km).
The figures do not include any signal decays in the Monte Carlo samples.
The background estimates, which give the dominant systematic uncertainty, 
are verified by comparing the numbers of events in the data and Monte 
Carlo samples in the $M > 1.8 \gevcc$ region.

The numbers of events selected in the data and Monte Carlo simulations are 
given in Table~\ref{table:kaon}.
The backgrounds predicted by the Monte Carlo simulations are approximately
equal to the numbers of events in the data sample.
There is an excess of data events in the \taukfourpipiz mode, but this
excess extends to mass values above the \mtau mass, indicating that 
events are due to background \mtau decays or \qqbar events.

The upper limits on the branching fractions are given in Table~\ref{table:kaon}.
There are no predictions for these modes.
We estimate that $\mathcal{B}(\taukfourpi) \sim 10^{-5} - 10^{-6}$ if the 
decay is related to $\mathcal{B}(\taufivepi)$ by the ratio of the 
CKM matrix elements ($V_{us}/V_{ud}$).
The \taufivepipiz decay is dominated by decays to the narrow low-lying 
resonances and the branching fraction of decay modes created by replacing 
a \pim with \Km would be highly suppressed due to the limited phase space.

\begin{table*}[tbph]
\renewcommand\arraystretch{1.25}
\begin{center}
\caption{\label{table:kaon} 
Results and branching fractions for charged kaon decay modes.}
\vspace{0.25cm}
\begin{tabular}{lccc} \hline \hline
& & & \\
Decay Mode            & \taukfourpi & \taukoppfourpi & \taukkppp \\
\hline 
Limit (90\% C.L.)     & $\kfournewlimit$  
                      & $\koppfourlimit$
                      & $\kkfourlimit$ \\
Branching fraction $(10^{-6})$    
                      & \hspace{0.5cm} $\kfourbrx$    \hspace{0.5cm}
                      & \hspace{0.5cm} $\koppfourbrx$ \hspace{0.5cm}
                      & \hspace{0.5cm} $\kkfourbrx$   \hspace{0.5cm} \\
Data events           & $\kfourdata$  
                      & $\koppfourdata$ 
                      & $\kkfourdata$ \\
Background            & $\kfourtotal$
                      & $\koppfourbc$
                      & $\kkfourbc$   \\
Selection Efficiency  & $\kfoureffx$
                      & $\kfoureffx$
                      & $\kkfoureffx$ \\
\hline
& & & \\
                      & \taukfourpipiz & \taukoppfourpipiz &  \\
\hline 
Limit (90\% C.L.)     & $\kfivelimit$  
                      & $\koppfivelimit$ 
                      &  \\
Branching fraction $(10^{-6})$    
                      & \hspace{0.5cm} $\kfivebrx$    \hspace{0.5cm}
                      & \hspace{0.5cm} $\koppfivebrx$ \hspace{0.5cm}
                      &  \\
Data events           & $\kfivedata$  
                      & $\koppfivedata$ 
                      &  \\
Background            & $\kfivetotal$
                      & $\koppfivebc$ 
                      &  \\
Selection Efficiency  & $\kfiveeffx$
                      & $\koppfiveeffx$ 
                      &  \\
\hline \hline
\end{tabular}
\end{center}
\end{table*}

\section {Summary}

We present measurements of the branching fractions for $\tau$
lepton decays to three-prong and five-prong final states.
The results are summarized in Table~\ref{table:summary}.
The branching fractions exclude contributions of the \KS meson.
The results are more precise than previous measurements and many decay
modes are studied for the first time.

\begin{acknowledgments}
We are grateful for the 
extraordinary contributions of our \pep2\ colleagues in
achieving the excellent luminosity and machine conditions
that have made this work possible.
The success of this project also relies critically on the 
expertise and dedication of the computing organizations that 
support \babar.
The collaborating institutions wish to thank 
SLAC for its support and the kind hospitality extended to them. 
This work is supported by the
US Department of Energy
and National Science Foundation, the
Natural Sciences and Engineering Research Council (Canada),
the Commissariat \`a l'Energie Atomique and
Institut National de Physique Nucl\'eaire et de Physique des Particules
(France), the
Bundesministerium f\"ur Bildung und Forschung and
Deutsche Forschungsgemeinschaft
(Germany), the
Istituto Nazionale di Fisica Nucleare (Italy),
the Foundation for Fundamental Research on Matter (The Netherlands),
the Research Council of Norway, the
Ministry of Education and Science of the Russian Federation, 
Ministerio de Ciencia e Innovaci\'on (Spain), and the
Science and Technology Facilities Council (United Kingdom).
Individuals have received support from 
the Marie-Curie IEF program (European Union) and the A. P. Sloan Foundation (USA). 
\end{acknowledgments}

\begin{center}
\begin{table*}[tbp!]
\renewcommand\arraystretch{1.25}
\caption{\label{table:summary}
Summary of branching fractions
excluding contributions from $\KS \rightarrow \pip\pim$.}
\vspace{0.25cm}
\begin{tabular}{lr}  
\hline\hline
Decay Mode   &  Branching fraction \\
\hline 
Resonant decays              &  \\
\taupppe (including $f_1$)  \hspace{4cm}
                             & $\sumetabavg$   \\
\taupppe (excluding $f_1$)     & $\sumetaexclusive$ \\
\taupzzeta (including $f_1$)  \hspace{0.25cm} 
                             & $\tauetaCorrBRPiTwoPiz$ \\
 & \\
\taufpi via \fpppp           & $\ffbrex$ \\
\taufpi via \feta            & $\sumtaufoneetabavg$ \\
$\mathcal{B}(\fpppp) / \mathcal{B}(f_1 \rightarrow \pi \pi \eta)$ 
                             & $\ratiofourptoppe$ \\
& \\
\taupppo                     & $\ombrb$ \\
\taupzzomega                 & $\BROmegaTwoPiPizCorr$ \\
\hline 
Non-resonant decays          & \\ 
\taufivepi (excluding $\omega$, $f_1$)     
                             & $\fpbf \times 10^{-4}$ \\
& \\
\taupppzzz (excluding $\eta$, $\omega$, $f_1$)    
                             & $\BRPSIncThreePiz$ \\
\taupppzzz (excluding $\eta$, $f_1$)              
                             & $\BRPSIncThreePizNoEtaorFone$ \\
 & \\
\taufivepipiz (excluding $\eta$, $\omega$, $f_1$) 
                             & $\sixbr$ \\
\taufivepipiz (excluding $\eta$, $f_1$)           
                             & $\sixbriso$ \\

\hline 
Inclusive decays (including  $\eta$, $\omega$, $f_1$)  & \\
\taupppzzz                   & $\BRTotalIncThreePiz$ \\
\taufivepi (excluding $\omega$)  & $\fpbfinc \times 10^{-4}$ \\
\taufivepipiz                & $\sixbrsum$ \\
\hline 
\multicolumn{2}{l}{\etapr (958) decays (90\% upper level confidence limit)} \\
\taupietaprimepiz            &  $< \pietaprimepizlimit$ \\
\tauketaprime                &  $< \ketaprimelimit$ \\
\tauetaprime                 &  $< \sumpietaprimelimitb$ \\
\hline
\multicolumn{2}{l}{Kaonic decays (90\% upper level confidence limit)} \\
\taukfourpi                & $< \kfournewlimit$    \\
\taukoppfourpi             & $< \koppfourlimit$ \\
\taukkppp                  & $< \kkfourlimit$   \\
\taukfourpipiz             & $< \kfivenewlimit$ \\
\taukoppfourpipiz          & $< \koppfivelimit$ \\
\hline \hline
\end{tabular}
\end{table*}
\end{center}



\begin{thebibliography}{99}
\bibitem{novo}
The Novosibirsk function is defined as \\
$f(m) = A \; {\rm exp} ( -0.5 \left\{ {\rm ln}^2 \left[ 1 + \Lambda \tau \cdot
(m - m_0) \right] / \tau^2 + \tau^2 \right\} )$,
where $\Lambda = \sinh(\tau \sqrt{{\rm ln}4}) / (\sigma \tau \sqrt{{\rm ln}4}
)$,
the peak position is $m_0$, the width is $\sigma$ and
$\tau$ is the tail parameter.
\bibitem{pdg:fone}  
$\mathcal{B}(\feta) = 0.349^{+0.013}_{-0.015}$ is obtained
by taking two-thirds of 
$\mathcal{B}(f_1 \rightarrow \pi \pi \eta ) = 0.524^{+0.019}_{-0.021}$
from Ref.~\cite{pdg}.
\end{thebibliography}
\end{document}